\documentclass{aa}  

\usepackage{graphicx}
\usepackage{gensymb}
\usepackage{algorithm}
\usepackage{algpseudocode}
\usepackage{txfonts}
\usepackage[colorlinks, citecolor=blue]{hyperref}

\usepackage{amsmath}
\usepackage{tablefootnote}

\usepackage[group-separator={,}]{siunitx}
\makeatletter
\renewcommand*\aa@pageof{, page \thepage{} of \pageref*{LastPage}}

\begin{document}

   \title{Mimicking the large-scale structure of the \\ Local Universe}

   \subtitle{Synthetic pre-labelled galaxies in large-scale structures}

   \author{M.~Alcázar-Laynez\inst{1}
          \and
          S.~Duarte Puertas\inst{1,2}
          \and
          S.~Verley\inst{1,2}
          \and
          G.~Blázquez-Calero\inst{3}
          \and
          A.~Jiménez\inst{1}
          \and
          A.~Lorenzo-Gutiérrez\inst{4}
          \and
          D.~Espada\inst{1,2}
          \and
          M.~Argudo-Fernández\inst{1,2}
          \and
          I.~Pérez\inst{1,2}
          }

   \institute{Departamento de Física Teórica y del Cosmos, Facultad de Ciencias
         (Edificio Mecenas), Universidad de Granada, E-18071 Granada, Spain
         \and Instituto Carlos I de Física Teórica y Computacional, Spain
         \and Instituto de Astrofísica de Andalucía, CSIC, Glorieta de la Astronomía s/n, E-18008 Granada, Spain.
         \and ATG Analytical, Avda de Andalucía 5, Granada, Spain}
              
	\date{Received June 26, 2025; accepted January 22, 2026}

\abstract
{Current observational and simulated large-scale structure (LSS) catalogues often lack consistency in assigning galaxies to specific structures, due to the absence of a universally accepted classification criterion.} 
{With the aim to generate synthetic empirical data for fine-tuning LSS
classification algorithms, as well as to train machine learning (ML) and deep learning (DL) models for the same purpose, this work presents a purely geometrical simulation
based on the statistical spatial properties found in LSS surveys, using the
spectroscopic main galaxy sample of the Sloan Digital Sky Survey (SDSS) catalogue up to a redshift of ${z\simeq0.1}$ as a specific use case.}
{A parallelism between the LSS and the 3D Voronoi tessellation was utilised,
in which the nodes, links, surfaces, and cells of the diagram correspond to
clusters, filaments, walls, and voids, respectively. The simulation used random
positions within voids as seeds for tessellating the 3D space.
The resulting tessellation structures were then randomly populated with galaxies
that adhere to the statistical properties of their observational respective structures. As the
galaxies were generated, they were tagged with their corresponding structure.} 
{In each simulation, six LSS mock catalogues were generated, following the statistical behaviour observed in the SDSS
catalogue, depending on the structure they belong to. In
addition, the Malmquist bias and the redshift-space distortion, known as the Fingers of God (FoG) effect,
were simulated as well.} 
{We present a novel geometrical LSS simulator, where generated galaxies mimic the
statistical properties of their observational belonging structure. As an example, the simulator was
tuned to mimic the SDSS catalogue, although any other catalogue can be considered in similar studies. With the generated catalogue, it is possible
to adjust the LSS classification algorithms, train and test ML and DL models, and
benchmark several LSS classification methods using this pre-labelled data to compare and
contrast their results and performance.}


\maketitle

\section{Introduction}
It is generally known that the environment of each galaxy affects its evolution.
In recent decades, studies have revealed the extensive arrangement of galaxies, consisting of
clusters, filaments, and walls, where galaxies are primarily
grouped together as a result of gravitational attraction \citep[][]{2020MNRAS.497..466S,
2024MNRAS.534.1682O, 2024PASA...41...96V}. In contrast, under-dense, sparsely populated areas known as voids exist as well. There have been limited opportunities to study these structures until large galaxy surveys started mapping the Universe
at large scales. The first attempt was the CfA Redshift Survey
\citep{1989Sci...246..897G}, which measured the radial velocities of
galaxies brighter than 14.5 magnitudes and using the Hubble-Lemaître law to obtain the first 3D reconstruction of part of the Local Universe.
Since then, numerous large-area surveys have been
performed, based on huge samples of galaxies  (e.g. Sloan Digital Sky Survey -SDSS-, \citealt{2000AJ....120.1579Y};
2dFGRS, \citealt{2001MNRAS.328.1039C}; VVDS, \citealt{2005A&A...439..845L};
GAMA, \citealt{2011MNRAS.413..971D}) and used to study these large-scale structures \citep[LSSs;][for SDSS, 2dFGRS, VIPERS, and GAMA, respectively]{2005ApJ...624..463G, 2004MNRAS.354L..61P, 2014A&A...566A.108G, 2014MNRAS.438..177A} 
and the
characteristics of galaxies within them. These studies have confirmed that such
structures appear both in the Local Universe and at greater distances, noting that the
structures resemble soap bubbles or  spider webs. However, the Universe is more
complex than this basic description and it might even present a hierarchical structure
\citep{2024MNRAS.527.4087J}.

Galaxies located in different structures present a wide range of characteristics. In an environment with nearby galaxies, which is common for clusters and filaments, gravitational forces work alongside other processes (e.g. ram pressure stripping and strangulation) to act as catalyst for star formation \citep[e.g.][]{1998ApJ...504L..75B, 2004MNRAS.348.1355B, 2002MNRAS.334..673L, 2005AJ....130.1482R}. Also, denser zones offer a hot intergalactic medium that also affects the evolution and properties of galaxies. Due to these interactions, their gas is forced to collapse and form
stars earlier than in galaxies located in voids or sparsely populated walls.
\citet{1980ApJ...236..351D} studied this gravitational effect in galaxies in
rich clusters, finding that the outer parts of the clusters are mostly composed of spiral galaxies with younger stellar populations than the elliptical and lenticular galaxies that tend to dominate in the inner parts of clusters.
To study the influence of the LSSs over their galactic
population, it is necessary to know the structure which these galaxies belong and
their typical characteristics: morphology, colour, stellar mass, size, amongst others. To date, several catalogues of galaxy clusters have been compiled
\citep[e.g.][]{1989ApJS...70....1A, 2012A&A...540A.106T,2014A&A...566A...1T,2017AA...602A.100T}, with  a number
of them based on SDSS data. For voids, we can refer to \citet{2011AJ....141....4K}
or \citet{2012MNRAS.421..926P}, and then to \citet{2014MNRAS.438.3465T} for background on filaments.
Currently, there are no observational catalogues of walls available.

In general, cluster galaxies and inner void galaxies can be relatively easy to identify, however, in some cases, there are discrepancies between the classifications determined by different authors. For instance, due to the lack of a commonly accepted definition of the LSS, some authors might classify a galaxy in the frontier of a void as a void galaxy, while others might classify it as a wall or filament galaxy.
There is no widely accepted and adopted criterion in the literature, which
means that several existing structure-specific algorithms end up in contradiction with one another in a way
that makes impossible to evaluate their consistency, when taken together or individually. For example, in \citet{2008MNRAS.386.2101N}, a parameter-free void-finding method was developed to find a differential property that would
allow for  a cosmological standard definition of voids to be established. However, if the
method is only focussed on a specific structure, we would have to use a range of different methods
for all the others, presenting a potential discrepancy between these methods. In
\citet{2018MNRAS.473.1195L}, 12 cosmic web structure classification methods are
visually and quantitatively compared to define an objective standard
rule. As a result, they found that each algorithm captures different properties of the cosmic
web, proving valuable in establishing the relationship between their results and
the effect on galaxies. Although some authors have used N-body simulators to mimic the
cosmic web and tune their algorithms, these simulations make no explicit
reference to the structure to which each galaxy belongs, making it impossible to
know exactly how reliable each method is. As explained in
\citet{2019MNRAS.484.5771A}, training data for semantic segmentation must satisfy characteristics, such as
establishing unique labelling systems that are consistent (i.e. the definitions of structures must be immutable across all data) and
diverse (i.e. sufficiently representative to capture the full spectrum of the
problem). This also applies to other ML-based methods and to mock testing data.

Due to this discrepancy, in this work, we present a geometrical LSS simulator that is
based on the spatial properties of the galaxies present in LSS catalogues,
according to their belonging structure, with well-defined sub-samples of galaxies
belonging to voids, walls and filaments, as well as clusters. Mock galaxies were
labelled at the time of their generation inside their corresponding structure, which
makes the catalogue an accurate reference for fine-tuning the LSS classification
algorithms, comparing and benchmarking their results against the empirical mock
catalogue, and training both machine learning (ML) and deep learning (DL) models to
find the defining features that allow us to differentiate between the galaxies of different
structures.

The structure of this article is organised as follows. Section~\ref{sec:2_Data}
presents the applied data, consisting of the catalogues we used as a reference to
extract the characteristics of each cosmic structure. In
Section~\ref{sec:3_Methodology}, we describe the initial conditions and methodology of the
simulation of LSS galaxies in the mock universe, while  the
characteristics of a generated mock universe using this method are presented in
Section~\ref{sec:4_Results}. Their quantitative and qualitative
features are inspected and compared with the reference catalogues in
Section~\ref{sec:5_Discussion}, along with several proposals for future
improvements. Finally, a summary of this work and the main conclusions of this
study are given in Section~\ref{sec:6_Conclusions}. In this work, a lambda cold dark matter ($\Lambda$CDM)
cosmology is assumed, with ${H_0\,=\,69.32}$\,[km\,s$^{-1}$\,Mpc$^{-1}$],
${\Omega_M\,=\,0.287,}$ and ${\Omega_{\Lambda}\,=\,0.713}$.

\section{Data: Observational catalogues}
\label{sec:2_Data}
As a practical application, in this work, the SDSS \citep{2015ApJS..219...12A} data were considered
as a general LSS catalogue which shape, biases and geometrical limits can be used as parameters to reproduce a synthetic observed universe in the
simulator.
In the present work, an indicator of the population and distribution of galaxies
within each structure is required. To that aim, two different
observational catalogues were used, namely, \citet{2012MNRAS.421..926P} and
\citet{2017AA...602A.100T}, as described in brief below:

\begin{itemize}
        \item \citet{2012MNRAS.421..926P}:  a public catalogue of
                cosmic voids based on SDSS Data Release 7
                \citep[SDSS-DR7,][]{2009ApJS..182..543A}. The voids were extracted using
                the Void Finder method developed by \citet{1996ApJ...462L..13E}. This
                catalogue contains \num{79947} galaxies distributed across
                \num{1055} voids.

    \item \citet{2017AA...602A.100T}:  to characterise
        clusters, the present work makes use of this galaxy groups catalogue,
        which contains \num{584449} galaxies belonging to \num{88662} groups.

\end{itemize}
There are several studies in the literature that investigate the distribution of
galaxies in SDSS according to their structure, which can be used as a reference
to set different values in the simulation configuration. For example, Fig. 8 of
\citet{2014MNRAS.441.2923C} presents a theoretical model estimating the volume
and mass distribution of galaxies in the SDSS catalogue as a function of the
structure in which they reside, according to the method applied by the authors.
\cite{2018MNRAS.473.1195L} presented a comparison between twelve different
methods to determine the volume and mass distribution in LSS, which will be
contrasted and compared against the resulting mock universes to
determine the goodness of the presented simulator in these values. Additionally,
the application LSSGalPy developed by \citet{2017PASP} was used to visualise the
LSS of the simulated mock catalogues in multiple space projections. 

\section{Methodology}
\label{sec:3_Methodology}
Observational LSS catalogues show the spiderweb morphology that the Universe
presents at large scales. There is a large number of geometric parameters that
can be extracted from these structures: length of filaments, area of the walls,
volume of voids, number of filaments intersecting in clusters, and others. In
the same way, it is possible to obtain certain properties: number of galaxies
per length, as well as the surface and volume units in filaments, walls, clusters and voids,
respectively.
Although there is no fixed number for these parameters, ideally, to create a mock universe only one single parameter is necessary: the volume of the Universe to be generated.
Any other parameter (e.g. the number of galaxies) must
be constant or only depend on this single parameter, presenting a statistical
behaviour that every mock universe must fit and reproduce to generate
the most realistic data. All the parameters of the presented simulator only depends of the desired simulation size and are estimated following stochastic models described below.

\subsection{3D mock universe parameters}
\label{subsec:2_1_3_3d_mock}
In the early Universe, mechanisms such as cosmic inflation established the initial
conditions of the universe, such as homogeneity and isotropy. The existing
irregularities in the early Universe resulted in a non-uniform mass distribution
that established the footprint of the current galaxy distribution within it while it expanded \citep{2007gitu.book.....S}. To generate mock
universes similar to the observed one, we can make use of Voronoi tessellations. The Voronoi tessellation can be calculated by growing (or expanding)
spheres radially from the positions of the generator points. The intersection of
two or more spheres establishes the edges, surfaces, and vertices of the Voronoi
tessellation for those generator points. With the Voronoi tessellation, it is
possible to recreate the shape of the LSS, considering edges as filaments,
surfaces as walls, vertices as galaxy clusters, and empty zones as voids. Using
this interpretation, it is possible to generate random points representing
galaxies along these structures, while controlling their density and location. In the literature, the
Voronoi tessellation is widely used by some authors to study the LSS. For
example, it was used by \citet{1997ApJ...491..421E} to test
void-finding algorithms populating a simulated universe with void and wall
galaxies. \citet{1987A&A...184...16I, 1989A&A...213....1V, 2007PhDT.......196A,
2010ApJ...723..364A} used it to interpret the LSS,
obtaining a consistent correlation with the observations. Furthermore, the
Voronoi tessellation not only serves as a simulator back end, but also as a
parameter estimation method. For instance, the Voronoi Tessellation was used in
\citet{2008MNRAS.386.2101N} to estimate the local densities of galaxies and
find void frontiers.

In the present simulator, filaments and walls are treated as the same structure
because: i) there are no filament catalogues complete enough to extract scaling relationships or intrinsic properties characterising these structures to implement them in the present simulator; and ii) at the time of writing, edges in the Voronoi tessellation
(scipy.spatial.Voronoi) from SciPy \citep{2020SciPy-NMeth} are not well indexed
for iterative processing. Specifically, For a given Voronoi cell, the edge
was occasionally listed twice (populated with galaxies twice, doubling the filament
density), while other times, an edge was never listed at all; as a result, no
galaxies were generated for those unlisted filaments.
Additionally, since this simulator produces a
geometrical representation of the LSS without specifying the intrinsic
properties of each galaxy, all simulated galaxies are generated with an unitary stellar
mass. Despite these considerations, to build the main blocks of the
simulator, it is necessary to characterise the following parameters:
\begin{itemize}
    \item  volume and shape of the generated mock universes;
        \item total count of structures: voids, clusters, walls and filaments;
        \item galaxy spatial distribution and density in each structure;
        \item know maximum and minimum sizes, areas, and volumes for each structure
        in the observed Universe.
\end{itemize}

As mentioned previously, in this work, the SDSS catalogue is used as the reference to
extract these parameters as a case study. We note that the SDSS occupies
${\sim}1/6$ of the complete sky. The simulator generates a whole-sky simulation
and then it is cut into six slices with the shape of the SDSS, generating six mock
universes at once. Thus, some parameters (such as the number of voids) will be
scaled up according to this difference. All adjusted parameters and their values
utilised for the SDSS use case are summarised in Table~\ref{table:default_config}.

\begin{figure}
    \begin{center}
        \includegraphics[width=0.79\hsize]{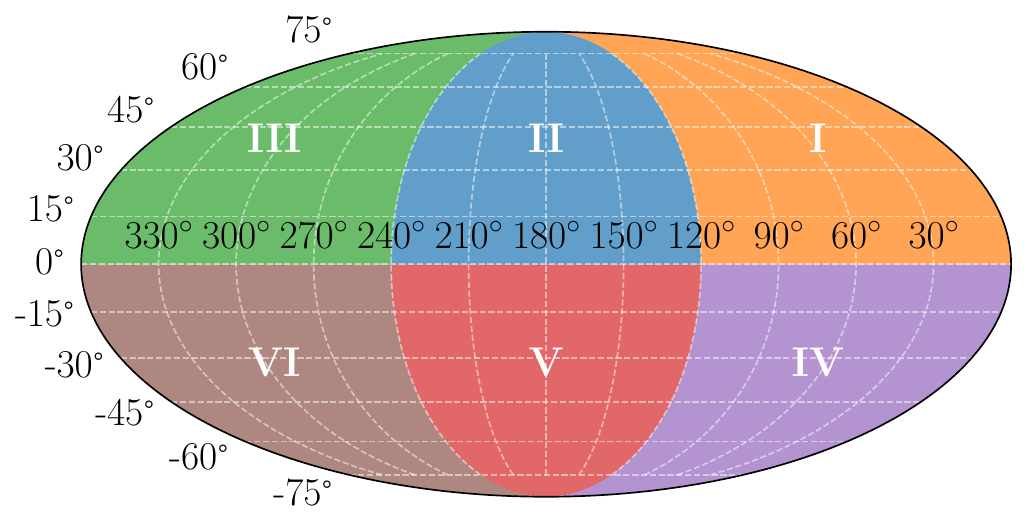}
            \caption[Spatial limits of each generated mock catalogue]{Spatial limits of each generated mock catalogue. Each slice, represented with different colours, ranges 120º in R.A. and 90º in Dec., and expanding radially from 0 to 500 [h\textsuperscript{-1}Mpc].} 
        \label{fig:sdss6slices}
    \end{center}
\end{figure}

The first parameter needed by the simulator is the size of the simulated
universe. To generate the mock universes with a shape  resembling the what is indicated by the
SDSS 3D spatial data, sphere-shaped universes with a radius of
500\,[h\textsuperscript{-1}Mpc] can be generated, corresponding approximately to a
redshift of $z \simeq 0.1$. The majority of the SDSS catalogue data occupy
from $120\degree$ to $240\degree$ in right ascension (R.A.) coordinates and from
$0\degree$ to $90\degree$ in Declination (Dec.) coordinates. Taking into account this
fact, this spherical whole-sky simulation is then sliced into six portions,
giving six simulated mock catalogues with the same shape and volume as the
reference one. In Fig.~\ref{fig:sdss6slices}, we show the spatial extent of each
slice.

The next parameter to determine is the number of voids to be
generated, which is done at random within the volume of the
simulation. \citet{2012MNRAS.421..926P} found approximately $\num{1000}$ voids in the SDSS catalogue. Scaling this parameter to the whole
sky, in the present simulation, a number of $6 \times \num{1000} = \num{6000}$ voids were set
by default. However, other authors such as \citet{2017ApJ...835..161M} estimates
the number of voids in catalogue around \num{1200}.
This
discrepancy can be explained because of the differences defining void limits, as well as the irregular shape of cosmic voids, where (depending
on the method used) the same void can be defined in sub-parts, increasing their
total number. For example, in \citet{2023ApJS..265....7D}, the number of
detected voids goes from \num{518} to \num{1184} depending on the method and cosmology used. Thus, in the present simulator, this parameter will vary between $6
\times [\num{500}, \num{2000}] = [\num{3000}, \num{12000}]$.

\subsubsection{Cluster galaxies density and distribution}

The parameters needed to generate the cluster galaxies are the statistical
distribution of the number of galaxies in a cluster, their spatial distribution,
and the distortion in redshift space caused by the dispersion in their radial
velocities due to the random peculiar velocities of galaxies within them, known
as the Fingers of God (FoG) effect. In the case of clusters, it is difficult to
determine the cluster galaxy density for several reasons. For instance, there
are faint galaxies that cannot be resolved optically or distinguished from other
kinds of object, leading to uncertainty when trying to define physical
boundaries of clusters and their shapes (let alone the spatial redshift
distortion). Nevertheless, for the purposes of this work, this parameter was
estimated to obtain an initial value, which could then be modified in case the
differences between the cluster simulation and the reference catalogues were, in
fact, clearly seen.

The typical stellar masses of clusters of galaxies are
${{\sim}10^{13}-10^{15}M_\odot}$ and their radii are of the order of 1-4\,Mpc.

Assuming that each galaxy has a mean stellar mass of  $\bar{M}_\star \simeq
10^{11} M_\odot $ and considering a typical cluster stellar mass of $M_{\rm
cluster} \simeq 10^{14}~M_\odot$, the  expected number of galaxies in a cluster
is $n=\num{1000}$ galaxies. Assuming a typical spherical cluster with radius
2~Mpc, the mean density, $\eta$, is expressed as

\begin{align*}
    \eta &= \frac{n}{4/3 \times \pi \times 2^3[\textrm{(h\textsuperscript{-1}Mpc)\textsuperscript{3}}]}=29.8\approx30\,[\textrm{galaxies}/\textrm{(h\textsuperscript{-1}Mpc)\textsuperscript{3}}].
\end{align*}

Although this number can be used, it might be meaningless, as due to gravitational
attraction, there are more galaxies in the inner radius than in the outer ones.
Additionally, considering spherical shapes for clusters, at a given radial
range, the volume is smaller in the inner zones than the outer ones. These two
effects make the spatial distribution highly inhomogeneous, as discussed
in Appendix~\ref{ex:cluster_shells}. In the present simulator, instead of just
assuming a uniformly distributed random galaxies within a simulated cluster
given a density, the number of galaxies for each cluster is chosen with a
user-selectable probability distribution. Then, this number is used to give the
synthetic cluster a maximum radius as a function of its number of galaxies,
also chosen randomly with a distribution specified by the user. For these two
parameters (number of galaxies and maximum radius), four kinds of distributions
can be selected: a Gaussian one, such as the Navarro-Frenk-White \citep[NFW;][]{1996ApJ...462..563N, 1997ApJ...490..493N} and Einasto \citep{1965TrAlm...5...87E} profiles, or a SDSS-like distribution that follows the
behaviour of the clusters found in the catalogue. By default, the SDSS-like
distribution is used in all parameters, but this can be changed in the
configuration file provided with the software.

\begin{figure}
    \begin{center}
        \includegraphics[width=0.9\hsize]{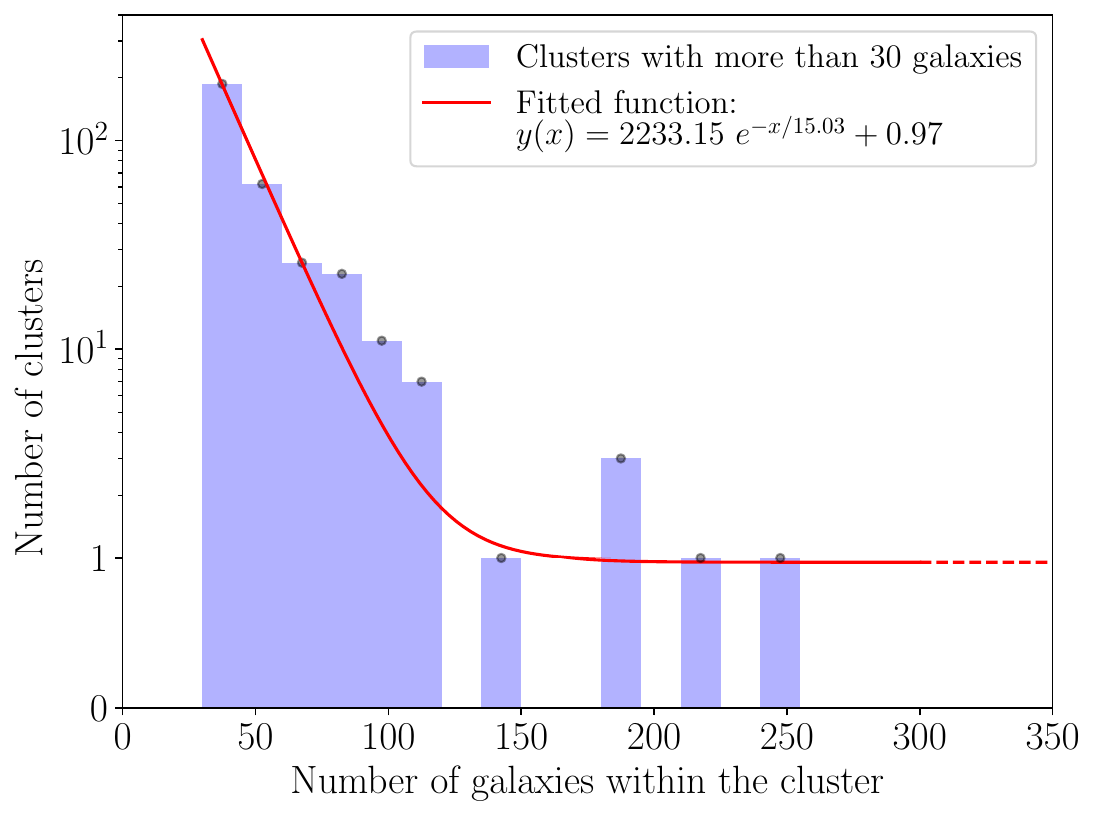}
            \caption[Number of galaxies per cluster]{Count of clusters as a
            function of their number of galaxies in the reference catalogue of
            \citet{2017AA...602A.100T}. Blue bars indicate the count of clusters. Red line indicates the fitted function over the data from the reference catalogue (used in
            the default configuration).}
        \label{fig:2_X_nGal_clusters}
    \end{center}
\end{figure}

\begin{figure}
    \begin{center}
        \includegraphics[width=0.9\hsize]{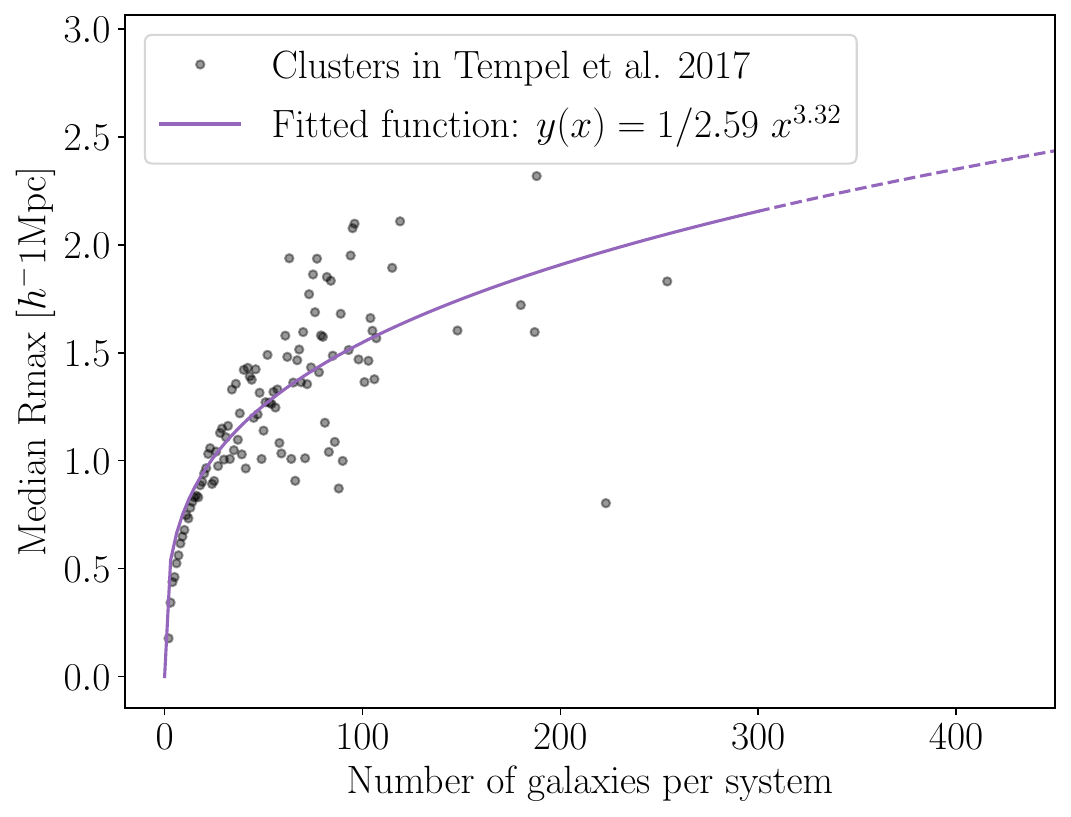}
        \caption[Rmax vs. N. of Galaxies]{Maximum radius of clusters as a 
        function of their number of galaxies in the reference catalogue.
        Black dots indicate the clusters from \citet{2017AA...602A.100T}. Purple line indicates the fitted function over the reference catalogue data.} 
        \label{fig:2_X_rmax_sdss_clusters}
    \end{center}
\end{figure}

As often found in the literature,  a group of galaxies with 30 or
more systems is considered a cluster \citep{1989ApJS...70....1A}. Also, in the
SDSS catalogue, the most populated cluster contains ${\sim}300$ galaxies. These two
indicators can guide us when we are trying to limit the number of galaxies in the synthetic clusters.
A prior study on the distribution of the number of cluster galaxies in SDSS was
carried out using data from \citet{2017AA...602A.100T}, resulting in the distribution shown in
Fig.~\ref{fig:2_X_nGal_clusters}. Considering the minimum and maximum number of
galaxies found in SDSS clusters, we can fit an exponential function to
build up a probability density function $P_{n_i,C_i}(x)$ to feed the synthetic
cluster generation code,

\begin{center}
    \begin{math}
        P_{n_i,C_i}(x) = \frac{1}{\num{4822.50}} (\num{2233.15} \times e^\frac{-x}{15.03} + 0.97);x \in [30, 300],
    \end{math}
\end{center}

\noindent where $C_i$ is the cluster number $i$ and $n_i$ is the number of
galaxies in that cluster. In addition to the number of galaxies, the radius of
clusters (i.e. the dispersion of the distance between a cluster galaxy and its
cluster geometric centre) can also follow either a Gaussian distribution or a
SDSS-like one. As shown in shown in Fig.~\ref{fig:2_X_rmax_sdss_clusters}, for
the SDSS-like distribution, a study of the distribution of the radius of
clusters as a function of their number of galaxies in SDSS was carried out using
data from \citet{2017AA...602A.100T}, following 

\begin{center}
    \begin{math}
        R_{max,C_i}(n_i) = \frac{1}{2.59} n_{i}^{1/3.32} \textrm{[h\textsuperscript{-1}Mpc].}
    \end{math}
\end{center}

\begin{figure}
    \begin{center}
        \includegraphics[width=0.9\hsize]{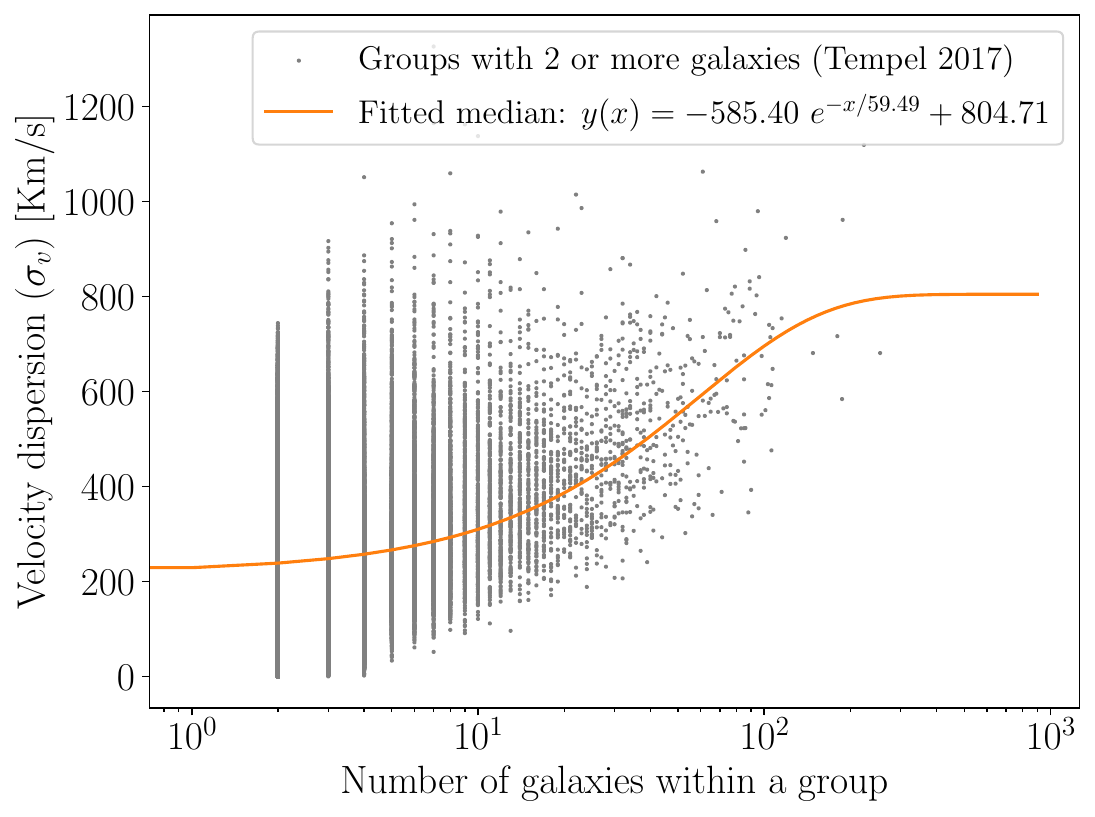}
        \caption[$\sigma_v$ vs. Number of galaxies in groups]{Fitted function
        for the FoG effect using \citet{2017AA...602A.100T} data.
        Black dots indicate groups in the reference catalogue. Orange line indicates the fitted function over the reference data.} 
        \label{fig:2_X_FoG_fitting_clusters}
    \end{center}
\end{figure}

\noindent Here, $C_i$ is the cluster number $i$ and $n_i$ is the number of galaxies in 
that cluster. Values below 0 or above 4\,[h\textsuperscript{-1}Mpc] are 
discarded and another random number is injected until the value reached between [0, 4].
The next parameter aims to recreate the radial velocity dispersion bias (also known
as the FoG effect) when observing physically bound dense areas.
This effect solely depends on the local density. A limited study on the
velocity dispersion in clusters as a function of their number of galaxies (i.e.
the local density) in SDSS was made using data from \citet{2017AA...602A.100T}  by
fitting the median of the velocity dispersion for each specific number of
galaxies in groups. The results, plotted in
Fig.~\ref{fig:2_X_FoG_fitting_clusters}, were obtained via

\begin{center}
    \begin{math}
        \sigma_{v,C_i}(n_i) = -585.40 \times e^{-n_{i}/59.49} + 804.71 \textrm{[km/s],}
    \end{math}
\end{center}

\noindent where $C_i$ is the cluster number $i$ and $n_i$ is the number of galaxies in 
that cluster. Once $\sigma_{v,C_i}(n_i)$ is computed, a random radial 
displacement is applied to each galaxy of cluster $C_i$ using a normal 
distribution with $\mu=0$ and $\sigma=\sigma_{v,C_i}(n_i)$.

\subsubsection{Wall and filament galaxies configuration}
These two structures need to be treated together: when two walls intersect, the
region where these structures cross is approximately equal to the sum of the densities
of each intersecting wall. This is why -- despite the fact that it is not possible to control the density of filaments and walls separately -- it is possible to adjust the densities of these
structures based on our knowledge of this effect and reach the densities of the filaments and
walls from the observed Universe. Despite the controversy around defining filaments
and walls, due to the fact that the number of galaxies in these two regions
represents approximately 80\% of the entire Universe, it is assumed that the wall
and filament structures are characterised by the same mean galaxy density than the Universe, which
correspond to approximately 0.1\,[galaxies\,$\times$\,(h\textsuperscript{-1}Mpc)\textsuperscript{-3}]. In the outer parts of the Voronoi
tessellation, larger cells are generated due to the lack of boundary generator
points and, thus, immense computer-intensive walls and filaments will be
populated. To dispose of them, a wall area limit can be programmed to avoid having to
populate walls and filaments above this limit. Taking all this into
account, the parameters needed to configure the wall and filament galaxy generation procedure are
as follows:

\begin{itemize}
        \item surface density of galaxies in walls per [h\textsuperscript{-1} Mpc]\textsuperscript{2};
        \item maximum area of walls in [h\textsuperscript{-1} Mpc]\textsuperscript{2};
        \item  distribution in the spatial separation of galaxies from the plane of their assigned wall;
    \item  parameters of the chosen distribution (mean, dispersion, ...).
\end{itemize}

The spatial separation parameter can be configured to follow a Gaussian or a
uniform distribution. The Gaussian distribution can be configured in mean and
dispersion, set with ${\mu=0}$\,[h\textsuperscript{-1}Mpc] and ${\sigma=0.5}$\,
[h\textsuperscript{-1}Mpc] as the default. For the uniform distribution, low and
high values of $-0.5$\,[h\textsuperscript{-1}Mpc] and
$0.5$\,[h\textsuperscript{-1}Mpc], respectively, were set. By default, it is
assumed that the spatial separation of galaxies from their wall plane
follows a normal (Gaussian) distribution.

\subsubsection{Void galaxies density and distribution}

In the present simulator, the user can select how to determine the number of
void galaxies for each simulation among these three options:
\begin{itemize}
        \item by density: generate a specified number of void galaxies per volume;
        \item by ratio: generate a specified number of void galaxies following
    the proportion with the number of galaxies from the other structures;
        \item fixed: specified number of void galaxies in the whole simulation, 
    regardless of any other parameter.
\end{itemize}

To establish a number of void galaxies using the density method in each simulation, \citet{2012MNRAS.421..926P} estimates a density of 0.01\,[galaxies\,$\times$\,(h\textsuperscript{-1}Mpc)\textsuperscript{-3}]. This number matches
the Calar Alto Void Integral-field Treasury surveY \citep[CAVITY,
][]{2024A&A...689A.213P} project results, obtaining approximately the same
density value for almost every studied void in its first public data release
\citep{2024A&A...691A.161G}.

Moreover, instead of using the density of voids, once galaxies in
other structures have been generated, we can estimate the number of void galaxies needed to follow the
mass proportions by structure. According to
\citet{2012MNRAS.421..926P}, it is known that void galaxies represent about the
10\% of the total of galaxies in the Universe. However, other works have reported different percentages. For example, following the theoretical models of
\citet{2007PhDT.......196A} and \citet{2014MNRAS.441.2923C}, this number can vary
from 15\% to 26\%. According to other theoretical results, such as those from the
studies compared in \citet{2018MNRAS.473.1195L}, the portion of void galaxies in
the LSS can vary from 10-15\% to >50\% depending
on the method, according to their simulations. By default, the simulator computes the number of void galaxies
using this last method (by ratio) with a default value of 15\% of void
galaxies. The final configurable parameter in void galaxies generation is the method used
to populate the space: i) populating the entire volume
with void galaxies (regardless if they are generated inside other structures,
thereby polluting them and inducing positional noise in the generated dataset); and ii) populating only
under-dense areas given a configurable threshold.

\subsection{3D mock universe generation}

We performed a 3D Voronoi tessellation procedure, where a specified number of random
points were spread between the given simulation spatial limits
using a uniform distribution. Once the tessellation was done, the different
structures were split and populated with galaxies independently.

The next step was to populate the generated regions with mock galaxies. Following
the Voronoi tessellation, in the mock model clusters can appear where two
or more filaments intersect (see Fig.~\ref{fig:2_1_4_cluster_vertex}), namely,
in tessellation nodes. Hence, for some nodes, random points (galaxies) are
generated and randomly distributed around them. The number of points follows the
distribution set in the configuration. The next algorithm shows how each cluster
is populated with galaxies:

\begin{algorithm}[h]
    \caption{Cluster galaxy generation algorithm}\label{alg:cap}
    \begin{algorithmic}
    \For{each cluster $C_i$}:
        \State Determine random number of galaxies $nOfGal$
        \State Determine maximum radius $rmax$ as a function of $nOfGal$
        \State Generate $nOfGal$ random points in a spherical distribution of radius $rmax$
        \State Translate the random points sphere to $C_i$ cluster centre 
        \State If selected, apply FoG effect by distorting the redshift of the galaxies
    \EndFor
    \end{algorithmic}
\end{algorithm}

\label{subsubsec:populating_walls_filaments}
Galaxies in walls and filaments have to be synthesised in four steps. The steps needed to generate the wall and filament galaxies are
as follows: i) obtain each cell in the tessellation; ii) extract each face of
the cell and, via an affine transformation, reduce the dimension of the face
to a 2D face; iii) add uniformly randomly disposed points in this 2D
face along its x and y coordinates, ensuring the points are inside the face, 
then apply a random z coordinate with a Gaussian distribution to each one;
and iv)  reverse the affine transformation made previously both for the
face and for every generated point, then update the mock dataset with this
new set of generated mock galaxies. This steps are schematised in the following
algorithm:

\begin{algorithm}[h]
    \caption{Wall and filament galaxies generation algorithm}\label{alg:wap}
    \begin{algorithmic}
    \For{each wall $W_i$};
        \State Reduce the dimension of $W_i$ to XY plane to get a flat 2D polygon.
        \State Compute area of wall $areaW_i$.
        \State Compute number of galaxies $nOfGal$ as a function of the configured density and $areaW_i$.
        \State Generate $nOfGal$ random points distributed in the flat 2D polygon.
        \State Give each point a random Z coordinate.
        \State Restore the 3D dimension of $W_i$ containing the random points.
    \EndFor
    \end{algorithmic}
\end{algorithm}

By repeating this algorithm for each face of a region, it is possible to obtain
the mock void with its enclosing walls and filaments galaxies, as shown in
Fig.~\ref{fig:6_wall}. The generation of void galaxies presents several
challenges for which design decisions have had to be made. It is possible to
just generate random points as mock void galaxies along the mock universe, no
matter which specific void each galaxy belongs to. However, if there is no
control over the coordinates where void galaxies are generated, they could grow
inside another region, thereby injecting categorisation errors into the mock
universe and into the processes that could be fed with it. Thus, a
density-driven method was developed to generate void galaxies in under-dense
areas exclusively, which is determined by a given threshold in the configuration
file. Once all the galaxies of the others structures have been generated, a 3D
histogram is performed, resulting in the count of galaxies per voxel. Then, the
histogram is inverted, where the voxels with the maximum count of galaxies will
contain the value of `0' and the voxels that previously had zero galaxies will
now contain the maximum value of the previous histogram. This inversion offers a
direct density map of the positions to generate the void galaxies. A graphical
example is shown in Fig~\ref{fig:inverted_3d_hist} representing the inverted
histogram extracted during the generation of a random mock universe. If the user
opts to populate the entire space with void galaxies (regardless of whether it
pollutes other structures), random points are created within the mock universe
using a uniform distribution. In the opposite case, the user can select to
populate only the under-dense areas, for which the following method is applied:

\begin{algorithm}[h]
\caption{Void galaxies generation algorithm}\label{alg:vap}
\begin{algorithmic}
    \State Generate a voxelised probability histogram of the current mock universe, with clusters and walls already populated.
    \State Invert the probability of each voxel by doing ${P_{Void}= 1-P_{Cluster,Wall}}$.
    \State Generate $nOfGal$ random points.
    \State Distribute the random points using the inverted probability histogram as a density map.
\end{algorithmic}
\end{algorithm}

In any case, the generation of void galaxies is uniformly distributed. Thus, a
constant density of void galaxies is expected, which does not depend on the distance to
the centre of voids \citep[in line with][]{2012MNRAS.421..926P}. The use of this method results in the galaxy spatial
distribution shown in Fig.~\ref{fig:2_1_5_void}, which is being simplified to
two dimensions for the sake of clarity. In the figure generated uniformly distributed void galaxies avoids to appear in dense areas. We can also clearly see how computing the void zones using voxels results in irregular wall fitting, presenting that distinctive triangular frontier near walls, which illustrates an
extreme staircase effect: voxelisation artifacts in computational geometry
and 3D modeling when a continuous shape, such as a sphere, is represented using
discrete elements such as voxels. Because voxels have a cubic shape and cannot
provide a perfect fit to curved surfaces, the process introduces volume errors and surface irregularities. As illustrated in
Fig.~\ref{fig:samplingvoxels}, the induced error decreases proportionally with
smaller voxel sizes.

\subsection{Mass and volume distribution of structures}
There are several characteristics that need to be measured in the mock
catalogues once all galaxies have been generated. These are used to analyse, compare, and ensure
agreement between these catalogues with the observational ones.
According to the stellar mass distribution, we assumed that galaxies
have a unitary mass. Thus, the total stellar mass of the mock universe is computed
as the sum of all the generated galaxies. The distribution of stellar mass is calculated
as the total stellar mass of the galaxies of a particular structure divided by the total
stellar mass of the mock universe, expressed as 

\begin{align}
M_{Total} &= M_{Cluster} + M_{Wall,Filament} + M_{Void};\quad \\
P_{Cluster} &= \frac{M_{Cluster}}{M_{Total}} \textrm{[\%]};\quad \\
P_{Wall,Filament} &= \frac{M_{Wall,Filament}}{M_{Total}} \textrm{[\%]};\quad \\
P_{Void} &= \frac{M_{Void}}{M_{Total}} \textrm{[\%]};\quad
\end{align}

\noindent where $M_{Total}$, $M_{Cluster}$, $M_{Wall,Filament}$, and $M_{Void}$
correspond to the total stellar mass of the simulation and the cumulative sum of the
clusters, the walls and filaments, and the voids stellar masses, respectively. Here,
$P_{Cluster}$, $P_{Wall,Filament}$, and $P_{Void}$ represent the portion of the
total simulated stellar mass represented by clusters, walls and filaments, and voids,
respectively.

Due to the difficulty in defining the LSS and their limits, the volume
estimation of the structures is not trivial. There exists several methods that
can be applied to a specific structure and then these results can
be compared and added to estimate the total volume. In the present simulator,
the following methods are implemented:

\paragraph{For clusters:}
\begin{itemize}
    \item Sigma methods: each cluster is assumed to occupy the volume of a sphere that contains 95.45\%, 98.76\% and 99.73\% ($2\sigma$, $2.5\sigma$, and $3\sigma$, respectively) of its galaxies.
    \item Convex hull method: the volume of each cluster is the same of a polyhedron envelope tangential to the most external galaxies, enclosing them all.
\end{itemize}

\paragraph{For walls and filaments:}
\begin{itemize}
    \item Sigma methods: each wall is assumed to occupy the volume of a polyhedron with the wall as its base and roof, parallel between them, and a given height that makes the polyhedron enclose the 95.45\%, 98.76\%, and 99.73\% ($2\sigma$, $2.5\sigma$ and $3\sigma$, respectively) of its galaxies.
    \item Convex hull method: the volume of each wall is the same of a polyhedron envelope tangential to the most external galaxies, enclosing them all.
\end{itemize}

\paragraph{For voids:}
\begin{itemize}
    \item Voxel method: the simulation volume is divided in voxels. For each
    voxel, if the count of galaxies of other structures is less than the
    mean density, the voxel is assumed to be a void one. The total void volume
    is the sum of all void voxels volume. This method is highly dependent on
    the voxel size due to the staircase effect.
    \item Indirect method: Knowing the total cumulative volume of clusters,
    and walls and filaments, the total void volume can be calculated
    by subtracting them from the total simulation volume as
    ${V_{Void} = V_{Total} - V_{Wall,Filament} + V_{Cluster}}$.
\end{itemize}

Sigma estimation methods are common in the literature (see
\citealt{2018MNRAS.473.1195L}). These operations present some caveats: different
methods could show disagreement in their results; for example, by interpreting a spatial
region as belonging to two structures, counting this space twice and therefore
overestimating the total volume. In the same way that this work aims to establish  a single
method to classify all structures, the volume estimation should also be computed
using a structure-independent single method. In the present work, the voxel method is
expanded to cover all structures: the simulated volume is divided into voxels.
For each voxel, if the count of galaxies of a specific structure is greater
than the count of the others, the voxel is assumed to belong to this
structure. If there are no galaxies within the voxel, it is added to the void structure.
The total volume of each structure is the sum of all voxels volume of a given
class. In case of a tie, the space of the voxel is divided by the number of tied
classes. As noted before, this method is highly dependent on the voxel size due
to the fact that (using the default configuration) the shape of the simulated
universe is spherical. When voxels are used to divide the volume of a sphere, sampling errors that are inversely proportional to the voxel size are present: the
smaller the voxel, the closer the sum of voxels volume is to the volume of the
sphere. This staircase effect is illustrated in Fig.~\ref{fig:samplingvoxels}
in the appendix.

\subsection{Generating the output dataset}
\label{subsec:2_1_3_5_generating_final_dataset}
Before the spherical simulated catalogue was divided into six parts (as detailed
before), the Malmquist bias was simulated and applied as well, where randomly
selected galaxies were discarded from the simulation as a function of the
distance. The aim here was to mimic the fact that there are faint galaxies difficult to detect
optically. The more distant, the more undetectable galaxies, thus the number of
detected galaxies decreases with the distance. This bias requires a reference, for
which (in this use case) the SDSS catalogue was utilised to follow its
distribution of galaxies along the distance to the observer, as represented in
Fig.~\ref{fig:Faint_galaxies_loss}. This effect was implemented as
follows. First, a histogram of the galaxy count per inverse-$h$ megaparsec along the distance to
the observer was created for both the reference and the mock catalogues. As
the observational catalogue only occupies ${\sim}1/6$ of the full sky, the
observational count was multiplied by 6. Then, randomly chosen galaxies are
discarded in mock universe for each $h^{-1}Mpc$ until both the observational and mock
galaxies counts at that distance were equal. Discarded galaxies are shown as
the value \texttt{`0'} in the \texttt{`selected'} column, while the observable
ones count with an \texttt{`1'} in this column. As an exception, the parameters
of cluster galaxies and their probability distributions were obtained from the
observable Universe; therefore, these parameters were measured after they
were affected by the Malmquist bias. Thus, cluster galaxies were excluded from
the random selection of galaxies to be discarded. To mitigate this, the
number of Voronoi nodes populated as clusters is decreased with the distance to the
observer in the same proportion, as done in \citet{2017AA...602A.100T}.

All the explained methods were implemented using \cite{python} v3.11 programming
language for the program logic, SciPy \citep{2020SciPy-NMeth} for generating the
Voronoi tessellation of the simulated space (scipy.spatial. Voronoi), Astropy
\citep{astropy:2013, astropy:2018, astropy:2022} for handling astronomical
coordinates (astropy.coordinates.SkyCoord), and Numpy \citep{harris2020array}
and Pandas \citep{reback2020pandas, mckinneyprocscipy2010} for structuring and
handling the large amounts of generated data. The preliminary versions of the
code before optimisations were tested using PROTEUS, the supercomputing centre
of Institute Carlos I in Granada, Spain. After optimisations, the simulator can
run on regular computers\footnote{The code of the presented simulator is open source and
publicly available at \url{https://gitlab.com/astrogal/mocklss}.}.

\section{Results}
\label{sec:4_Results}
We generated several random mock universes. Each one lasted about twenty minutes
using a laptop with 64GB of RAM and a Intel i7 CPU with 16 logic cores. All
realisations were produced using the same procedure to ensure they would be
statistically consistent with one another. Thus, for the sake of
clarity, the results of a single randomly chosen mock universe and its six
slices (see
Fig.~\ref{fig:sdss6slices}) are featured. A representation of slice II is shown
in Fig.~\ref{fig:LSSGalPy_mock}. All results were measured after applying
Malmquist bias. The symbol "$\pm$" represents the standard deviation of
measured values (1$\sigma$)

The studied mock catalogue ranges from 0º to 120º, 120º to 240º, and 240º to
360º in R.A. and from -90º to 0º and 0º to 90º in Dec. The
farthest galaxy is ${{\sim}500}$\,[h\textsuperscript{-1}Mpc]. The total volume of
the mock universe is
\num{523598776}\,[h\textsuperscript{-1}Mpc]\textsuperscript{3}, where
each slice occupies
\num{87266463}\,[h\textsuperscript{-1}Mpc]\textsuperscript{3}. The
number of galaxies of each slice are ${\num{498943}\pm\num{5142}}$. The
mean density of galaxies in the studied mock is
${\eta_{mock}=mass_{mock}/volume_{mock}={\sim}~0.0057}$\,[galaxies~$\times$~(h\textsuperscript{-1}Mpc)\textsuperscript{-3}].
On average, the mass distribution of galaxies depending on their structure are
organised as follows: $3.29\%$ of galaxies belongs to clusters, $80.80\%$ of
galaxies belongs to walls and filaments, and the remaining $15.91\%$ of galaxies
belongs to voids, as illustrated in Fig.~\ref{fig:Mass_distribution}. The volume
distribution result as follows: $0.07\%$ of the volume of the mock universe
belongs to clusters, $19.33\%$ of the volume belongs to walls and filaments,
and the remaining $80.60\%$ of the volume belongs to voids. The Malmquist bias
was simulated as well, with the result presented in
Fig.~\ref{fig:Faint_galaxies_loss}, where galaxies follow the detection curve
present in the reference catalogue.

\begin{figure}
        \begin{center}
        \includegraphics[angle=0.0, width=0.9\hsize]{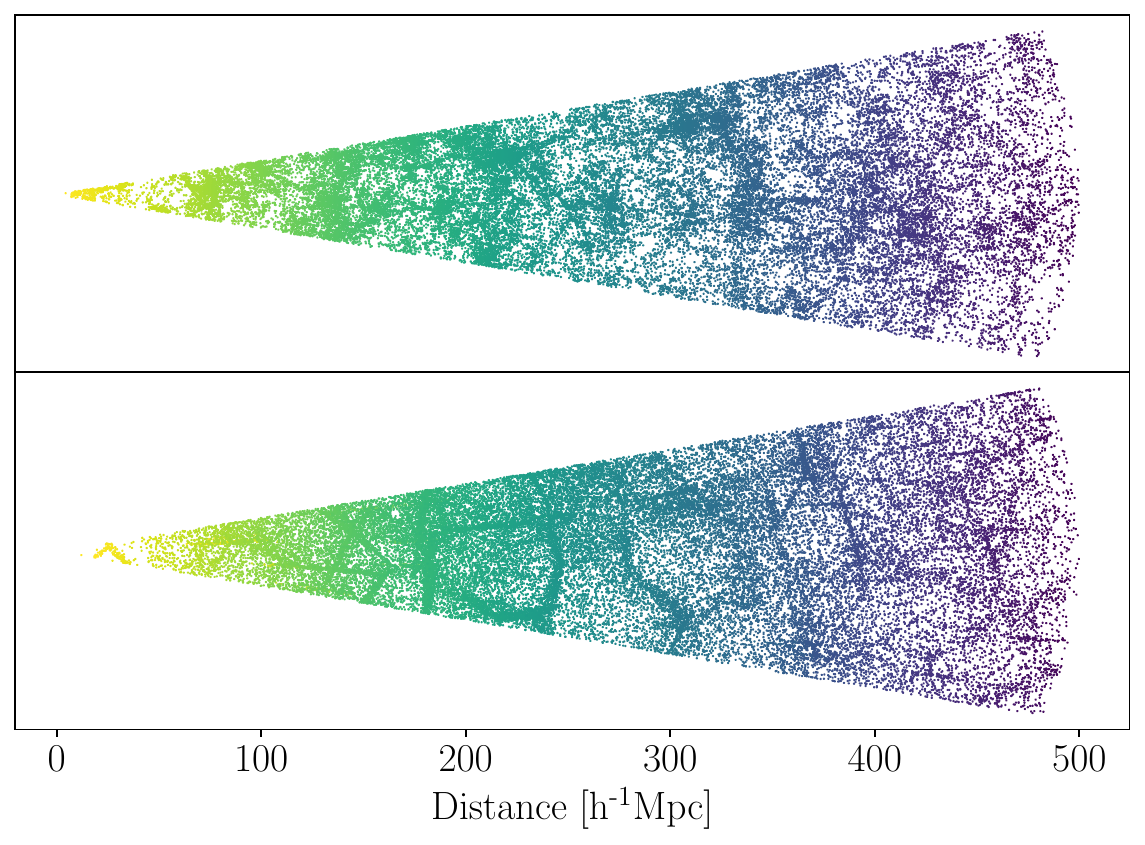}       
                \caption[Lightcone]{Light cone comparison between a 15[\textsuperscript{o}]
                sections of the reference SDSS catalogue \citep{2015ApJS..219...12A}
                in the top panel and the analysed mock catalogue in the lower panel. Colour
                represents the distance to observer (the darker the purple shade, the farther the distance).}
                \label{fig:Lightcone}
        \end{center}
\end{figure}

\begin{figure*}
        \begin{center}
        \includegraphics[angle=0.0, width=0.95\hsize]{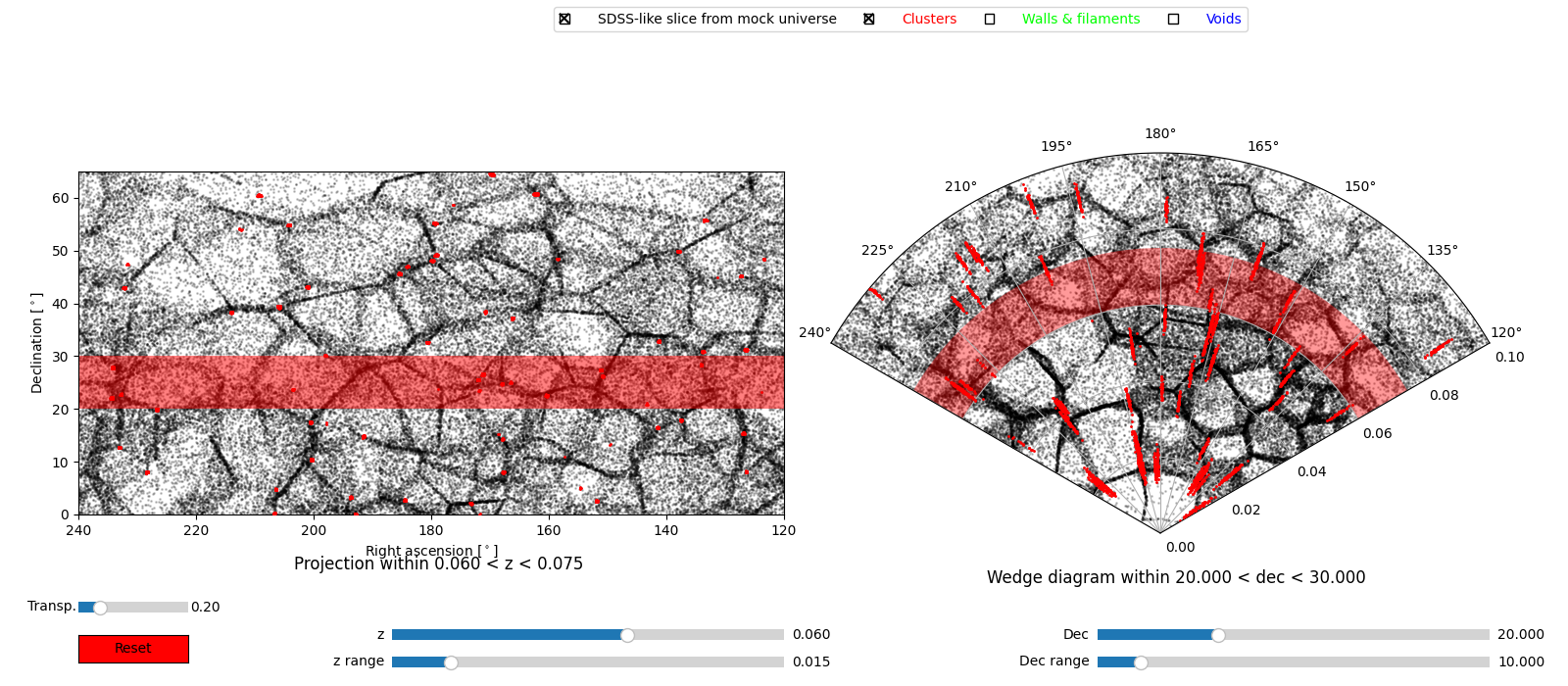}
        \includegraphics[angle=0.0, width=0.95\hsize]{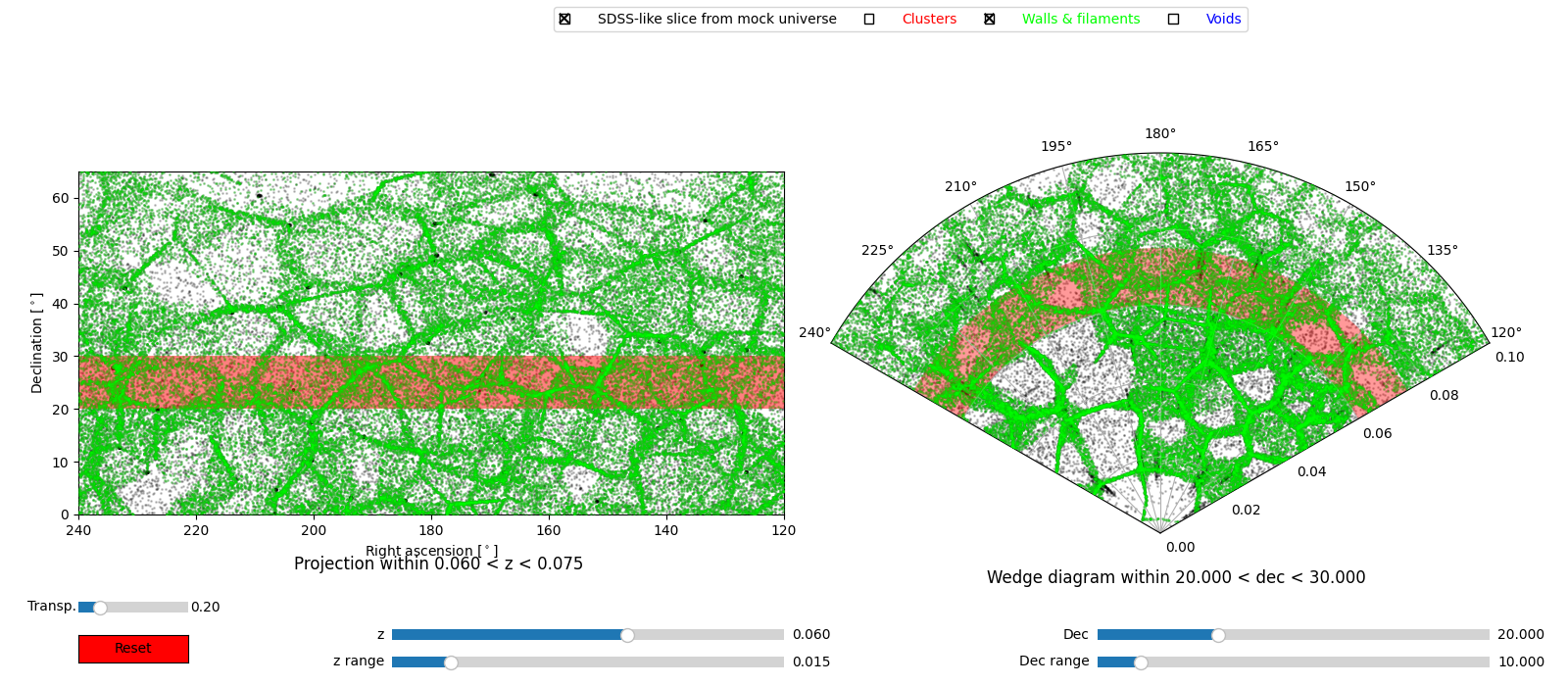}
        \includegraphics[angle=0.0, width=0.95\hsize]{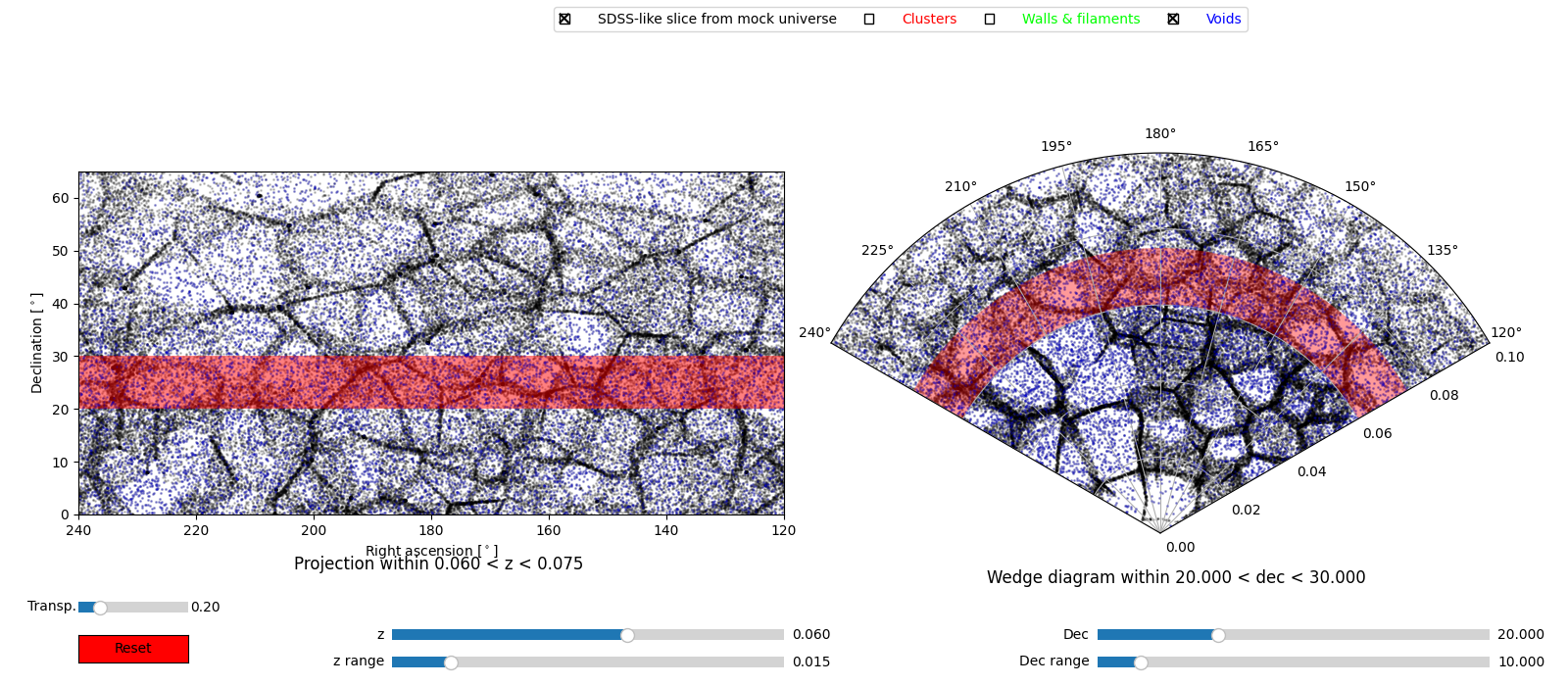}
                \caption[Mock SDSS-like slice]{SDSS-like slice II from a randomly generated mock universe. From top to bottom: Mock catalogue with cluster, wall and filament, and void galaxies coloured in red, green, and blue, respectively. In all panels, the red bands represent the range of redshift and declination shown.  
  Upper panel: Mock catalogue with cluster galaxies coloured in red. Middle panel: Mock catalogue with wall and filament galaxies coloured in green. Lower panel: Mock catalogue with void galaxies coloured in blue.}
                \label{fig:LSSGalPy_mock}
        \end{center}
\end{figure*}

\begin{figure}
        \begin{center}
        \includegraphics[angle=0.0, width=0.49\hsize]{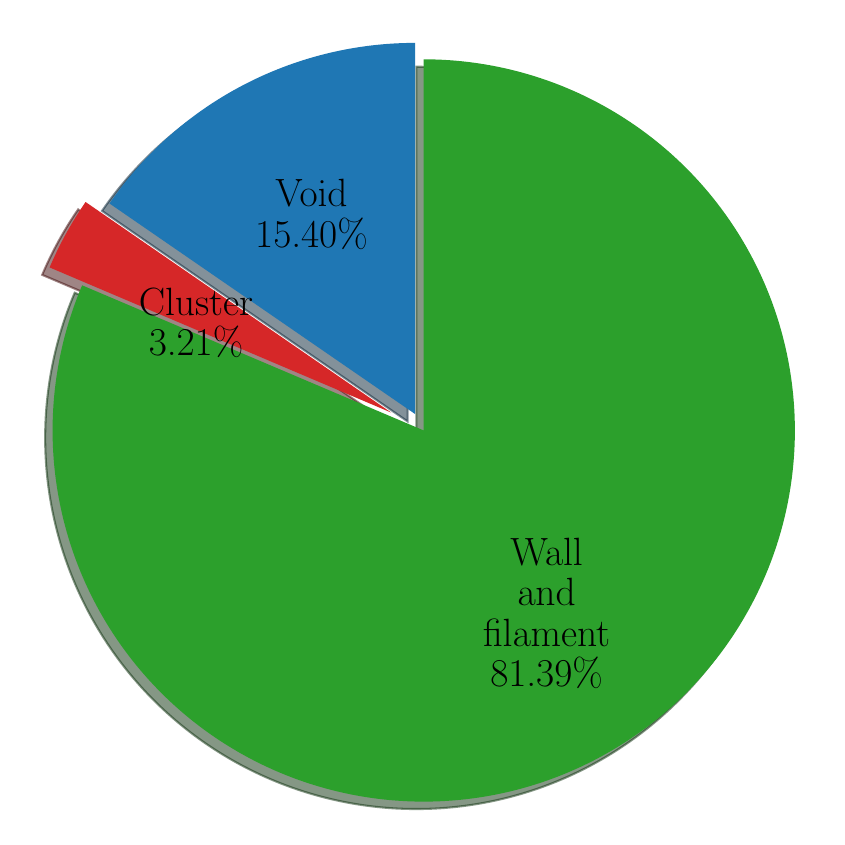}
                \includegraphics[angle=0.0, width=0.49\hsize]{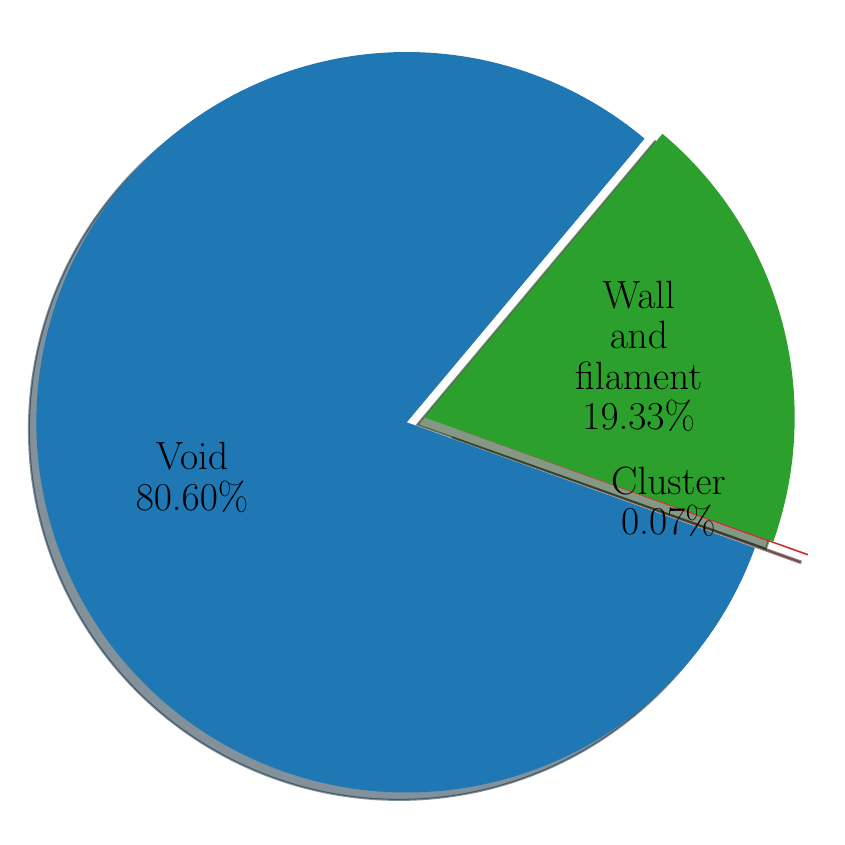}
                \caption[Mass (left panel) and volume (right panel) distribution per structure]{Mass (left panel) and volume (right panel) distribution of galaxies per structure in the studied mock universe. The colour determines the structure as in Fig.~\ref{fig:LSSGalPy_mock}.}
                \label{fig:Mass_distribution}
        \end{center}
\end{figure}

\begin{figure}
        \begin{center}
                \includegraphics[angle=0.0, width=0.95\hsize]{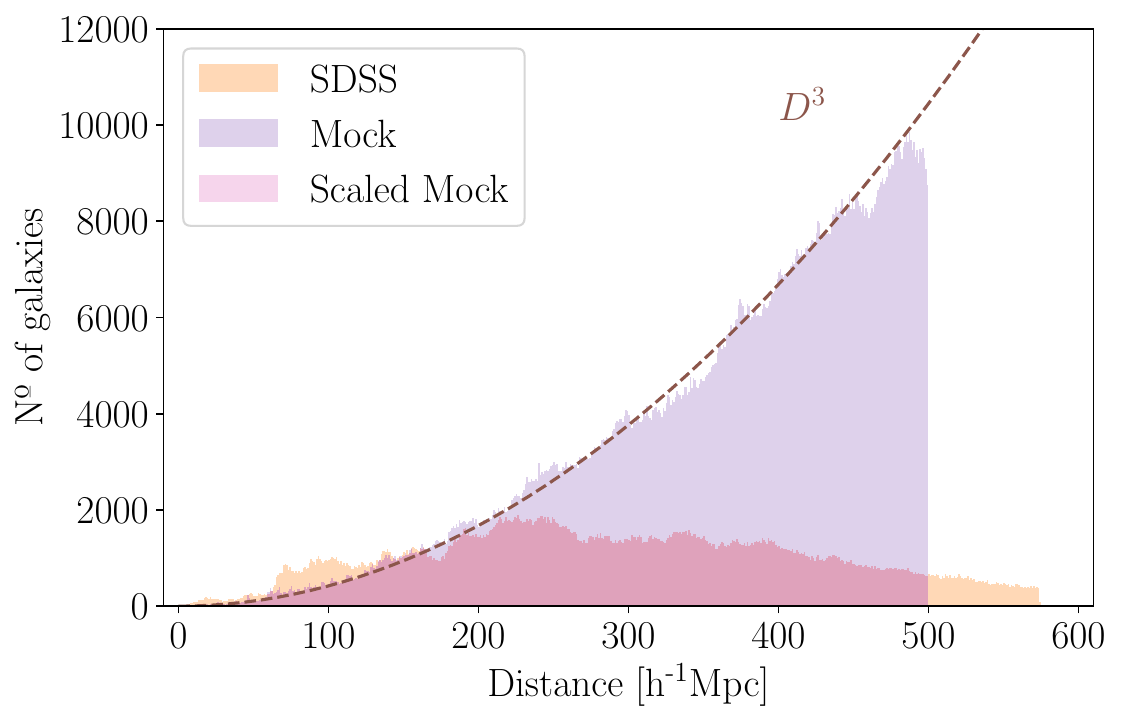}
                \caption[Malmquist bias simulation]{Simulated Malmquist bias. Orange
                histogram shows the distribution of the number of sampled galaxies in the
                observational reference catalogue along the distance from origin. Purple
                histogram shows the distribution of the number of generated mock galaxies in the
                mock catalogue along the distance from origin. Pink histogram shows the distribution of the number of the selected mock galaxies in the mock
                catalogue along the distance from origin, after discarding some mock
                galaxies randomly to follow the behaviour of the observational
                catalogue. Brown dashed line gives the expected number of galaxies for a constant
                volume density of galaxies within the universe.}
                \label{fig:Faint_galaxies_loss}
        \end{center}
\end{figure}

\subsection{Mock cluster properties}

\begin{figure}
        \begin{center}
                \includegraphics[angle=0.0, width=0.9\hsize]{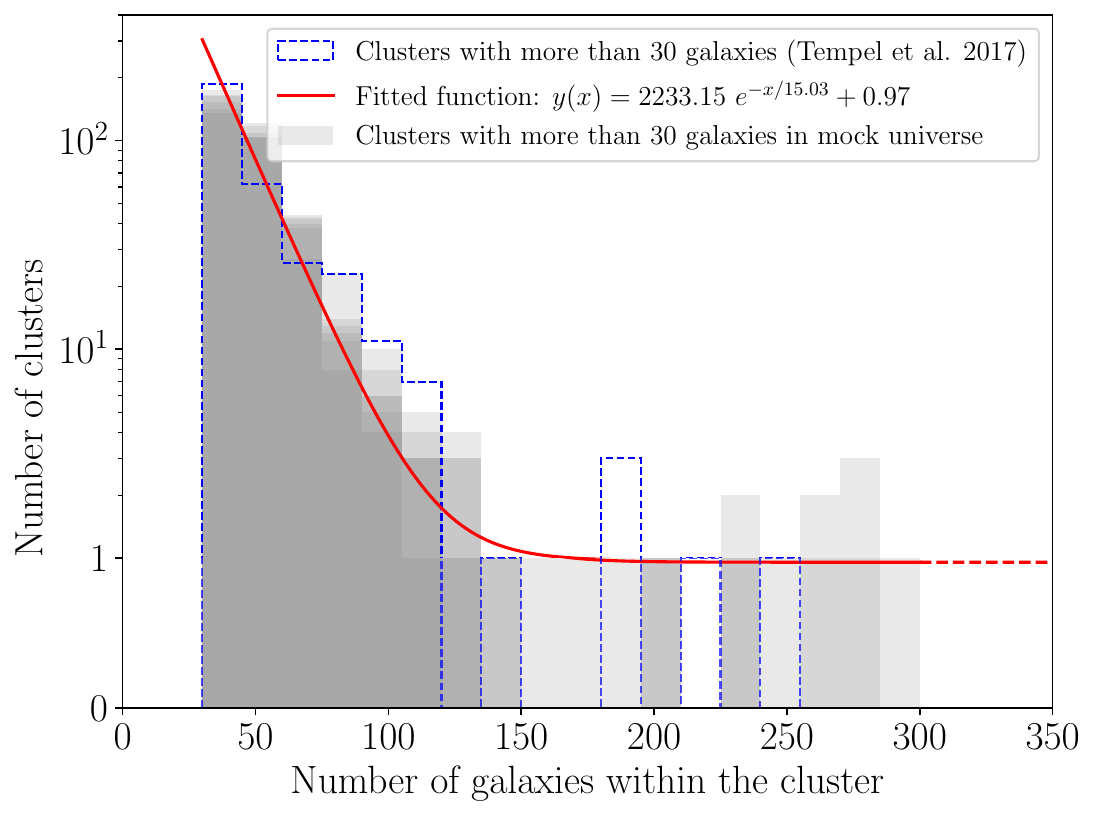}
                \caption[Number of galaxies in cluster]{Count of clusters as a function
                of their number of galaxies in \citet{2017AA...602A.100T} and in the
                simulated mock universe. Dashed blue line shows the number of clusters in
                \citet{2017AA...602A.100T}. Red line shows the fitted function over the data from
                the reference catalogue. Grey bars show the number of clusters in each slice of
                the studied mock catalogue.}
                \label{fig:Ngal_cluster_histogram}
        \end{center}
\end{figure}

\begin{figure}
        \begin{center}
                \includegraphics[angle=0.0, width=0.9\hsize]{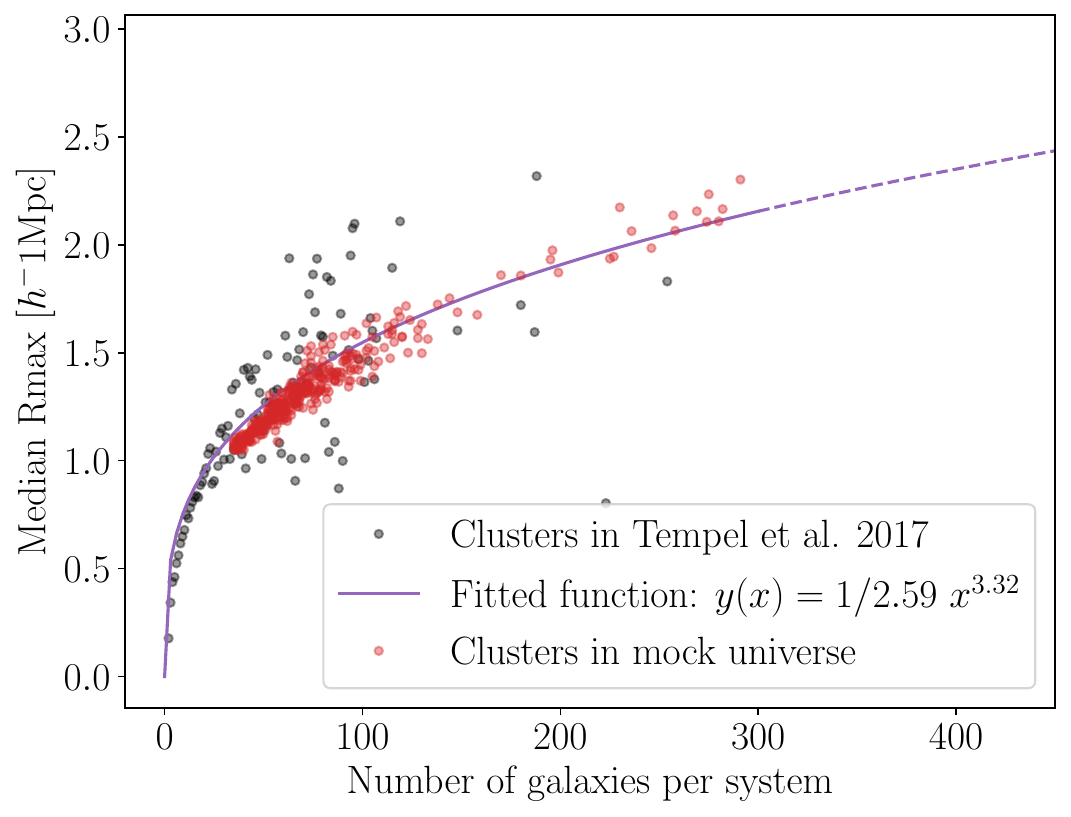}
                \caption[Rmax vs. N. of Galaxies in Tempel and mock]{Comparison between
                the maximum radius of clusters as a function of their number of galaxies
                in \citet{2017AA...602A.100T} (black dots) and in all slices of the
                generated mock universe (red dots).}
                \label{fig:rmax_vs_ngal_cluster}
        \end{center}
\end{figure}

\begin{figure}
        \begin{center}
                \includegraphics[angle=0.0, width=0.9\hsize]{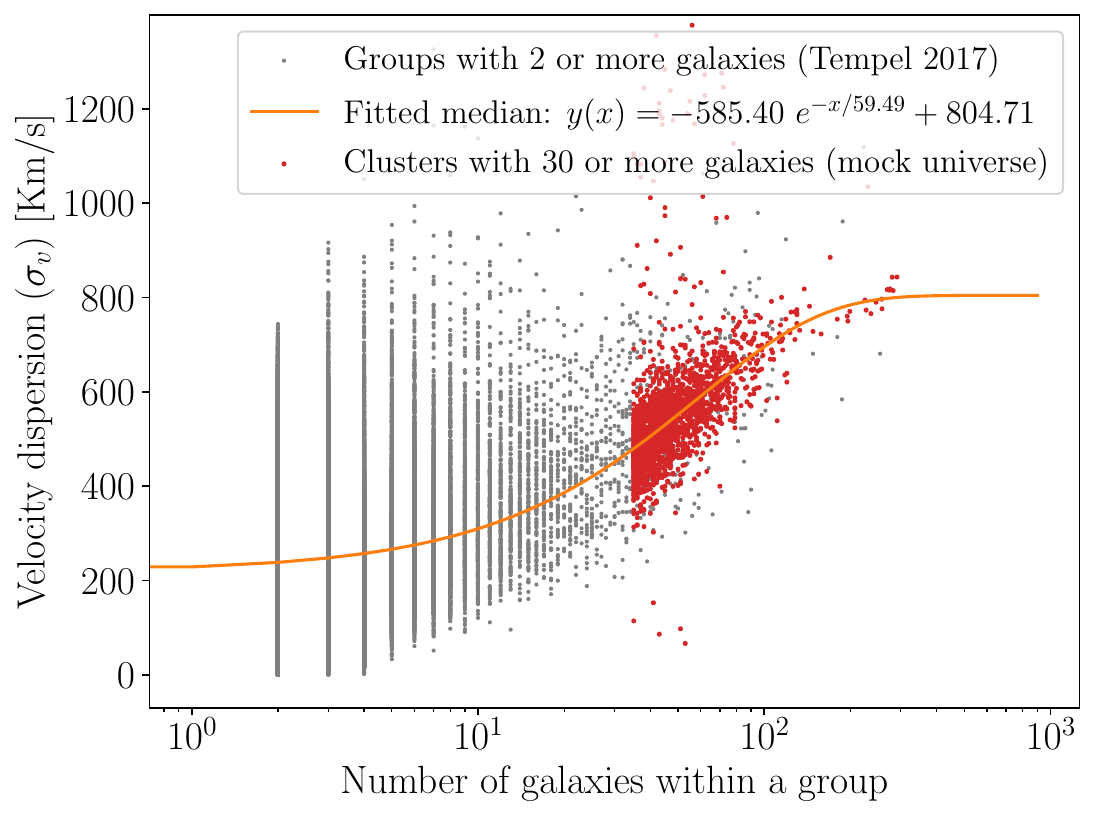}
                \caption[Fingers of God in Tempel and mock]{Comparison between the
                radial velocity dispersion bias as a function of the number of galaxies
                in groups in \citet{2017AA...602A.100T} (black dots) and in all slices of
                the generated mock universe (red dots), which follows the fitted 
                function over the reference catalogue data (orange line).}
                \label{fig:FoG_mock}
        \end{center}
\end{figure}

Each slice of the studied mock catalogue has $\num{16406}\pm\num{832}$ cluster
galaxies distributed in $\num{333}\pm\num{17}$ clusters, where each one of them contains 30
or more galaxies. The distribution in the number of galaxies within clusters is
shown in Fig.~\ref{fig:Ngal_cluster_histogram}. Depending on this number, each
cluster presents differences in their radii (as seen in
Fig.~\ref{fig:rmax_vs_ngal_cluster}) and in their velocity dispersion (see Fig.~\ref{fig:FoG_mock}). The mean density of galaxies is ${\eta_c = 0.267\pm0.014}$\,[galaxies~$\times$~
(h\textsuperscript{-1}Mpc)\textsuperscript{-3}]. The most populated cluster
contains $291$ galaxies. The number of connected filaments per cluster is four in all cases.

\subsection{Mock wall and filament properties} 
The slices of the generated mock catalogue contains $\num{403150}\pm\num{7643}$ wall and
filament galaxies distributed in $\num{15022}\pm\num{569}$ walls or filaments. As walls
and filaments represents $19.33\%$ of the total volume of the simulation, the
total volume of wall and filaments in the studied mock universe is about
$\num{16868607}$\,[h\textsuperscript{-1}Mpc]\textsuperscript{3} in each slice. The mean density
of galaxies in walls and filaments has the value of ${\eta_{w\&f} =
0.024}\pm0.0001$\,[galaxies\,$\times$\,(h\textsuperscript{-1}Mpc)\textsuperscript{-3}]. The most populated wall has \num{358}
galaxies.

\subsection{Mock void properties}
Each slice of the generated mock catalogue has
${\num{79386}\pm\num{3272}}$ void galaxies distributed in
${\num{1060}\pm\num{37}}$ cosmic voids. Voids represent the
80.60\% of the total volume of the simulation, making a total volume of
$\num{70336769}$\,[h\textsuperscript{-1}Mpc]\textsuperscript{3} in each
mock universe slice. The mean density of galaxies within void has the value of
${\eta_v =
0.001\pm0.00005}$\,[galaxies~$\times$~(h\textsuperscript{-1}Mpc)\textsuperscript{-3}].
This represents one-sixth of the mean density of galaxies in the mock universe.

\section{Discussion}
\label{sec:5_Discussion}

In this section, the characteristics of an arbitrary random mock universe are
compared with reference catalogues. The results can vary for other simulated
mock universes that use a different random seed, but their values are similar to
those shown in this section. These studied features are summarised in
Table~\ref{table:mock_analysis}.

A visual inspection allows us to check that the simulated galaxies are labelled
with their corresponding structure (as illustrated for slice II within
R.A.\,${\in[120º, 240º]}$, Dec.\,${\in[0º, 90º]}$ in
Fig.~\ref{fig:LSSGalPy_mock}). Also, Fig.~\ref{fig:Lightcone} shows the
filamentary structure of the simulation that is also given in the SDSS catalogue.
Once inspected, the following statistical analyses are performed and contrasted with
the reference catalogues. The total number of galaxies in the studied slices is
$\num{498943}\pm\num{5142}$. Within a footprint, common for both simulated
and observational catalogues ($RA\in[140^{o}, 230^{o})$, $DEC\in[0^{o}, 50^{o})$
and $Dist.\in[100, 500)[h^{-1}Mpc]$), the mock universe presents
$\num{275487}\pm\num{6700}$ galaxies, which is close to the reference
SDSS catalogue, containing $\num{274975}$ galaxies.

The mass distribution of galaxies depending on their structure
are organised as follows: $3.29\%$ of galaxies belong to clusters, $80.80\%$ of
galaxies belong to walls and filaments, and the remaining $15.91\%$ of galaxies
belong to voids, as illustrated in Fig.~\ref{fig:Mass_distribution}. These mass
and volume proportions are consistent with the results from other studies, such
as \citet{2007PhDT.......196A} and \citet{2014MNRAS.441.2923C}. In
\citet{2018MNRAS.473.1195L}, the algorithms of these and other authors are
compared. Comparing the mock universe proportions with these studies, in
general, this mock universe generator is able to keep these proportions when
replicating the reference observational catalogue. However, the overall mass in
clusters is lower than the estimations over the observational catalogue
(discussion in Sect.~\ref{sub:cluster_properties}). The Malmquist bias was also simulated, with the result presented in
Fig.~\ref{fig:Faint_galaxies_loss}. When the number of galaxies in the mock
universe is greater than in the reference catalogue for a given distance,
randomly chosen ones gets discarded until the detection curve of both catalogues
match. In this simulation, this occurs at ${\sim}$150\,[h\textsuperscript{-1}Mpc].

\begin{figure}
        \begin{center}
                \includegraphics[angle=0.0, width=0.83\hsize]{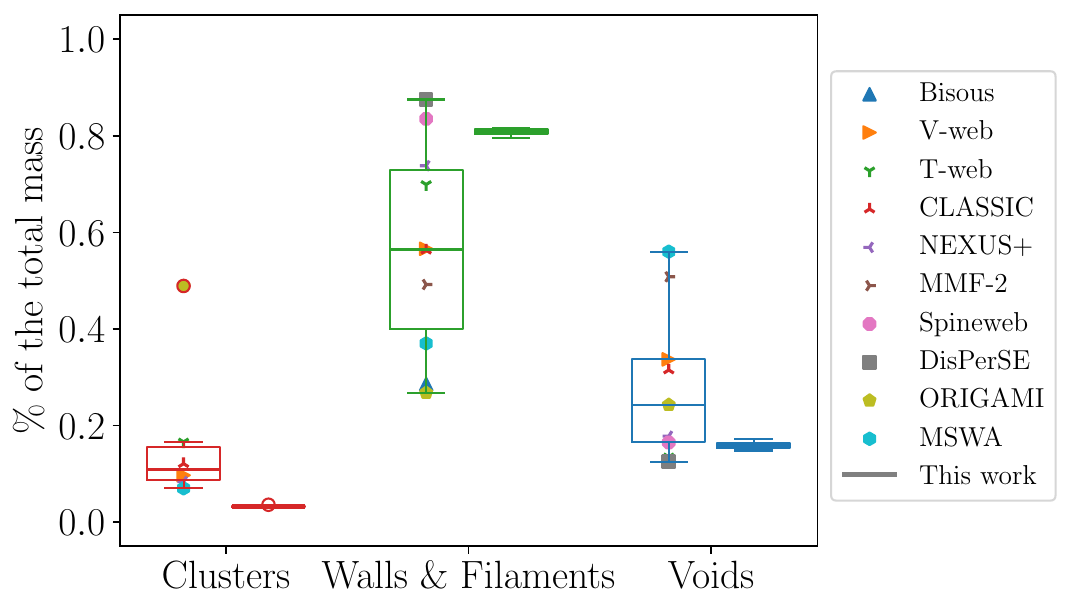}
        \includegraphics[angle=0.0, width=0.83\hsize]{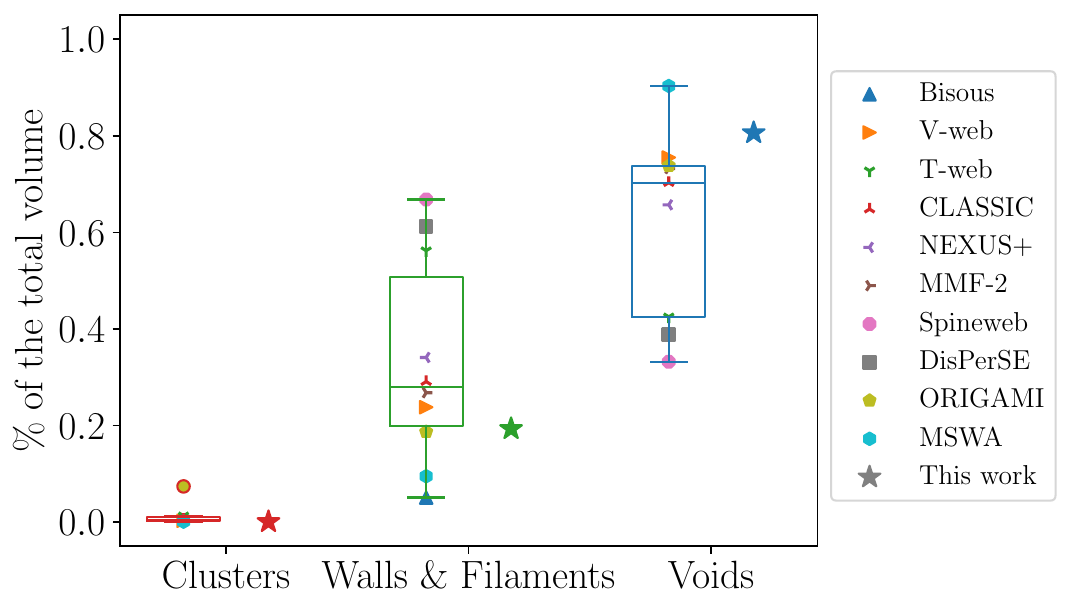}
                \caption[Distributions per structure against bibliography]{Mass (upper panel) and
                volume (lower panel) distribution of galaxies depending on their belonging structure,
                comparing the data of the analysed mock universe against the results of
                the methods compared in \citet{2018MNRAS.473.1195L} over observational
                SDSS data.}
                \label{fig:mock_vs_libeskind}
        \end{center}
\end{figure}

The features that we ought to measure next are specific to each structure. Due to the
lack of wall and filament catalogues, it is not possible to contrast
the characteristics in these two structures against observational data. In the
following sub-sections, the properties of each specific mock structures are
discussed.

\subsection{Cluster properties}
\label{sub:cluster_properties}
The studied mock catalogue has $\num{16406}\pm\num{832}$ cluster galaxies per
slice distributed throughout $\num{333}\pm\num{17}$ clusters, where each of them contains 30 or
more galaxies. These values are close to those obtained in the catalogue of
\citet{2017AA...602A.100T}, which contains $\num{16419}$ cluster
galaxies distributed throughout 322 clusters. The fraction of cluster galaxies in
the mock catalogue is slightly lower than that obtained with the methods
presented in \citet{2018MNRAS.473.1195L}. This occurs for several reasons.
First, the simulator produces galaxies with unitary mass, thus there are no
galaxies within clusters with more mass than in other structures. Second, the
number of clusters in the studied mock catalogue can deviate from the imposed
distribution (see Fig.~\ref{fig:Ngal_cluster_histogram}). This is a purely
random effect: in other simulations, it can be greater than the fitted function.
Third, this simulation is programmed to follow \citet{2017AA...602A.100T}
catalogue, which essentially has a smaller number of cluster galaxies than that of
\citet{2018MNRAS.473.1195L}. Finally, the simulator generates clusters, namely,
groups of galaxies with more than 30 galaxies. The methods presented in
\citet{2018MNRAS.473.1195L} detect knots, which do not necessarily contain 30 or
more galaxies. Because the simulation uses an isotropic Poisson-Voronoi 3D
tessellation, the number of connected filaments per cluster is equal to the
vertex degree of the tessellation ($4$), which matches the expected behaviour.

As seen in Fig.~\ref{fig:Ngal_cluster_histogram}, the mock universe simulator
is in line with the probability curve of the reference cluster catalogue when we are selecting a
random number of galaxies for populating each one. In addition, depending on the
selected number of galaxies, a maximum radius for each cluster is determined,
which also follows the statistical behaviour found in the observational
catalogue, as shown in Fig.~\ref{fig:rmax_vs_ngal_cluster}. The FoG effect was recreated successfully in the mock universe, where the cluster
galaxies present a radial velocity dispersion that is dependent on the number of
galaxies (a proxy of the total stellar mass) within their belonging cluster (as seen in
Fig.~\ref{fig:FoG_mock}), following the fitted curve shown in
Fig.~\ref{fig:2_X_FoG_fitting_clusters}.

\subsection{Void properties}
The generated mock catalogue has ${\num{79387}\pm\num{3272}}$
void galaxies distributed across ${\num{1060}\pm\num{37}}$ cosmic voids
per each slice. These numbers are consistent with those presented in
\citet{2012MNRAS.421..926P}, where \num{79947} void
galaxies distributed in \num{1054} voids were reported in the observational SDSS catalogue. In
this reference void catalogue, the mean density of galaxies in voids represents
the 10\% of the mean density of the SDSS catalogue, rising from 20\% to 100\%
near the borders of voids. In this simulation, the mean density of galaxies in
voids is about ${0.0010\pm0.00001}$\,
[galaxies~$\times$~(h\textsuperscript{-1}Mpc)\textsuperscript{-3}], which
represents the ${\sim18\%}$ of the mean density of the studied mock universe and
match the results of the reference void catalogue. Moreover, the density of the
mock void galaxies is uniform and not changing by the distance to the void
centre, which replicates the observed behaviour of the void galaxies from the
SDSS catalogue.

\subsection{Wall and filament properties}
The mean walls and filaments galaxy density is about four times the mean
density of the whole mock catalogue, which is far from that of the clusters and voids
($\num{4700}\%$ and ${18\%}$ of the mean density). This is expected when the
distribution of mass per structure is taken into account because wall and
filament galaxies represent approximately the 80\% of the total mass of both
mock and observational catalogues. Thus, it is expected that the mean density of
the wall and filament would be about the same order of magnitude as the mean total
density. Compared to \citet{2018MNRAS.473.1195L}, the wall and filament mass
fractions are expected to be above the mean, since galaxies in groups that are
not classified as clusters fall into this category.

Observational catalogues of filaments exist in the literature. However,
observational catalogues of walls are not yet available and so, walls are currently defined  from a primarily theoretical perspective (e.g.
\cite{2025arXiv251002419H}). In our framework, filaments are not explicitly
identified and they tend to emerge, specifically, at the intersections of two or more
walls. As a result, walls and filaments are treated as a single structural
category (i.e walls and filaments), for which no direct observational counterpart
currently exists.

\subsection{Impact and possible improvements}
This simulator presents several advantages, compared with other methods. First,
it allows us to handle key spatial characteristics of each structure, such as the
density of galaxies within each structure and its spatial position, allowing for the
construction of standard mock catalogues or even the generation of edge-case
scenarios by changing parameters to extreme values (e.g. the impact
of changing the number of voids to unrealistic values, as shown in
Fig.~\ref{fig:mock_100_10000}). Second, each galaxy has a unique single label,
meaning that the classification has a unique solution; thus, allowing us to avoid the
possibility of finding multiple minima for one class. Third, the definition of
the structures is consistent across the dataset, regardless of the position and
orientation of each structure, the distance to the observer, or the Malmquist
bias, thereby improving the generalisation of models. Finally, multiple numbers of
different universes can be generated by changing the random seed, which is
decisive for increasing the generalisation ability of ML models and, thus,
decreasing the overfitting.

Given these advantages, several LSS analysis methods would benefit from
the flexibility of the present simulator. For instance,
\cite{2019MNRAS.484.5771A} generates a Voronoi tessellation to represent the LSS
and performs its classification. However, instead of populating each structure
with galaxies, the space is voxelised and then entered to the convolutional
neural network (CNN). This makes the method highly dependent on the chosen voxel
size, imposing a classification based on the galaxies found within each
classified voxel, regardless of whether they belong to different structures (block
classification). Using these mock universes, the first layers of the CNN can be
redesigned, making it able to process each galaxy independently, and assigning a
structure to each one, regardless of the voxel size or other imposed spatial
parameter.

Besides these benefits, there are several improvements that can be
adopted in the present work  to generate more realistic datasets. For
example, walls and filaments are generated in a way that makes them extremely straight; whereas in the SDSS
catalogue, these structures clearly have a more organic form, with
irregularly shaped walls and curvy filaments. These shapes can be achieved by
expanding and shrinking the 3D space in the centres of voids,
randomly. This action can give random concave and convex shapes to both walls
and filaments. In addition, this simulator cannot generate wall and filament
galaxies independently. The implementation of this feature is crucial to
encourage the implementation of wall and filament galaxy classification
algorithms using this simulator as reference to measure their accuracies.
Nevertheless, this simulator is now ready to be used against the extensive list
of existing LSS classification algorithms, taking as an example, those presented
in \citet{2018MNRAS.473.1195L}.

\section{Summary and conclusions}
\label{sec:6_Conclusions}

We present a novel geometrical mock galaxy catalogue generator, where its
simulated galaxies follow the statistical behaviour of the observed ones in each
respective LSS of the Universe in which they reside. The proposed simulator
enables a precise customisation of structural properties, unambiguous labelling,
and consistency under varying observational conditions. In addition, the
variability introduced by random seeds enhances data generalisation (crucial to
reduce overfitting in ML and DL models). As a use case, the simulator was
fine-tuned to mimic the galaxies found in the SDSS catalogue. To this end,
several observational studies characterising the galactic properties based on
structure have been taken into account, such as \citet{2012MNRAS.421..926P} for
voids and \citet{2017AA...602A.100T} for clusters. It is not only the purely
structural properties that are included in the simulation, but also the
observational biases such as the FoG effect and the Malmquist bias are taken
into account. Among other parameters, these characteristics are fully
customisable. In this specific use case, the following results were obtained:

\begin{itemize}
    \item In each simulation, a full-sky mock catalogue was generated,
    which was then sliced into six parts where each one spread 120º in R.A.  and 90º in Dec., equivalent to the
    area covered by the SDSS footprint.

    \item Given the standard configuration, each slice of the mock catalogues presents a
    total number of ${\sim}\num{500000}$ galaxies, distributed on the basis of their
    affiliated structure, where ${\sim}\num{3}\%$ of galaxies belong to
    clusters, ${\sim}81\%$ of galaxies belong to walls and filaments, and
    ${\sim}\num{16}\%$ belong to voids. In terms of volume, the clusters, walls
    and filaments, and voids represent $0.07\%$, ${\sim}\num{19}\%$, and
    ${\sim}\num{81}\%$ of the whole simulation volume, respectively.

    \item The slices of the studied mock catalogue has ${\num{16406}\pm\num{832}}$,
    ${\num{403150}\pm\num{7643}}$ and ${\num{79387}\pm\num{3272}}$ galaxies distributed in
    ${\num{333}\pm\num{17}}$, ${\num{15022}\pm\num{569}}$ and ${\num{1060}\pm\num{37}}$ clusters, walls or filaments and cosmic
    voids, respectively.

    \item These results are consistent with those observed in the SDSS
    catalogue \citep{2012MNRAS.421..926P, 2017AA...602A.100T}, which is expected 
    due to the construction of the mock universe.
\end{itemize}

In conclusion, this simulator provides a robust and expandable platform
for building customisable mock universe catalogues. It enables a systematic
testing of LSS classification methodologies, supports the development of
ML approaches for galaxy surveys, and facilitates controlled
explorations of boundary-case scenarios. These mock catalogues can be used to
train ML models and evaluate their accuracy, which will be the subject of a future
work.

\section*{Data availability}
The source code of the presented simulator is open source and publicly available
at: \url{https://gitlab.com/astrogal/mocklss}.

\begin{acknowledgements}
      We dedicate this work to the memory of our dearest colleague and friend J. Jiménez-Vicente.
      The authors thank the anonymous referee for the thorough reading and constructive feedback.
      The CAVITY project acknowledges financial support by the research
      projects AYA2017-84897-P, PID2020-113689GB-I00, PID2020-114414GB-I00, and
      PID2023-149578NB-I00 funded by the Spanish Ministry of Science and Innovation
      (MCIN/AEI/10.13039/501100011033) and by FEDER/UE; the project A-FQM-510-UGR20
      funded by FEDER/Junta de Andalucía-Consejería de Transformación Económica,
      Industria, Conocimiento y Universidades/Proyecto; by the grants P20-00334 and FQM108,
      funded by Junta de Andalucía; and by Consejería de Universidad, Investigación e
      Innovación (Junta de Andalucía) and Gobierno de Espa\~{n}a and European Union
      NextGenerationEU through grant AST22\_4.4.
      M.A-F. acknowledges support from the Emergia program (EMERGIA20-38888) from Consejería de Transformación Económica, Industria, Conocimiento y Universidades of the Junta de Andalucía.
      Funding for SDSS-III has been provided by the Alfred P. Sloan Foundation, the Participating Institutions, the National Science Foundation, and the U.S. Department of Energy Office of Science. SDSS-III is managed by the Astrophysical Research Consortium for the Participating Institutions of the SDSS-III Collaboration including the University of Arizona, the Brazilian Participation Group, Brookhaven National Laboratory, Carnegie Mellon University, University of Florida, the French Participation Group, the German Participation Group, Harvard University, the Instituto de Astrofísica de Canarias, the Michigan State/Notre Dame/ JINA Participation Group, Johns Hopkins University, Lawrence Berkeley National Laboratory, Max Planck Institute for Astrophysics, Max Planck Institute for Extraterrestrial Physics, New Mexico State University, New York University, Ohio State University, Pennsylvania State University, University of Portsmouth, Princeton University, the Spanish Participation Group, University of Tokyo, University of Utah, Vanderbilt University, University of Virginia, University of Washington, and Yale University. The SDSS-III web site is {\tt \href{http://www.sdss3.org/}{http://www.sdss3.org}}.
      We are also grateful for the computing resources and related technical support provided by PROTEUS, the supercomputing centre of Institute Carlos I in Granada, Spain.
      This research made use of \cite{python} v3.11 programming language.
      This work made use of Astropy (\url{http://www.astropy.org}) a community-
      developed core Python package and an ecosystem of tools and resources for 
      astronomy \citep{astropy:2013, astropy:2018, astropy:2022}; SciPy \citep{2020SciPy-NMeth}; Numpy
      \citep{harris2020array}; and Pandas \citep{reback2020pandas, 
      mckinneyprocscipy2010}.
\end{acknowledgements}

\bibliographystyle{aa}
\bibliography{refs}

\clearpage
\appendix
\onecolumn

\section{Default simulator parameters (SDSS-like)}

\begin{table}[h]
\caption{Parameters considered in this work for the SDSS-like simulation.}
\label{table:default_config}
\centering
\begin{tabular}{lll}
\hline\hline
Default simulator configuration & & \\
Parameter & Value & Unit\\\\
\hline\hline
General configuration & &\\\hline
Seed & Random number based on hour and date & \\
Number of voids & \num{8089} & Voids\\
Maximum diameter of simulated universe  & \num{1000} & $\textrm{[h\textsuperscript{-1}Mpc]}$\\
Number of CPU cores & Maximum of the host machine & \\
Malmquist bias & True & Boolean \\
Reference catalogue & \citet{2015ApJS..219...12A} & \\
Grid size for volume calculation & 5 & $\textrm{[h\textsuperscript{-1}Mpc]}^3$  \\\\\hline\hline
Clusters &  \\

\hline
FoG bias & True & Boolean \\
Number of galaxies prob. distribution  & SDSS \\
Number of galaxies SDSS prob. distribution  & $P_{n_i,C_i}(x) = (\num{2233.15} \times e^{-x/15.03} + 0.97)/\num{4822.5}$ \\
Minimum number of galaxies per cluster & 30 & galaxies \\
Maximum number of galaxies per cluster & 300 & galaxies \\
Maximum radius prob. distribution & SDSS (fixed, depending on \# of gal. in cluster: $n_{i}$)& \\
Maximum radius SDSS formula & $R_{max,C_i}(n_i) = n_{i}^{1/3.32}/2.59$ & $\textrm{[h\textsuperscript{-1}Mpc]}$ \\
Minimum radius & 1 & $\textrm{[h\textsuperscript{-1}Mpc]}$ \\
Maximum radius & 4 & $\textrm{[h\textsuperscript{-1}Mpc]}$ \\
Radial velocity dispersion & $\sigma_v = -585.40e^{-n_i/59.49} + 804.71$ & km/s \\\\\hline\hline
Walls \& filaments &   \\

\hline
Wall galaxy density & {0.18} & galaxies $\times$ (h\textsuperscript{-1}Mpc)\textsuperscript{-2} \\
Minimum wall surface &    0 & $\textrm{[h\textsuperscript{-1}Mpc]}^2$ \\
Maximum wall surface & \num{2000} & $\textrm{[h\textsuperscript{-1}Mpc]}^2$ \\
Galaxy spatial distribution within a wall & Gaussian & \\
Galaxy spatial distribution mean & $\mu = 0$ & $\textrm{[h\textsuperscript{-1}Mpc]}$\\ 
Galaxy spatial distribution dispersion & $\sigma = 0.5$ & $\textrm{[h\textsuperscript{-1}Mpc]}$\\\\\hline\hline
Voids &   \\

\hline
Grid size for low density zones detection & 5 & $\textrm{[h\textsuperscript{-1}Mpc]}^3$  \\
Determination of number of galaxies & By ratio (depending on other structures) & \\
Number of galaxies ratio & $N_{void\,gal.} = {0.116} \times (N_{cl.\,gal.} + N_{w.\&f.\,gal.}) $ & galaxies\\\hline
\hline
\end{tabular}
\end{table}

\clearpage

\section{Results of a simulation given the default configuration}

\begin{table}[h]
\caption{Analysis of the slices of the studied mock universe.\tablefootnote{The symbol "$\pm$" represents the standard
deviation of measured values (1$\sigma$)}}
\label{table:mock_analysis}
\centering
\begin{tabular}{llll}
\hline\hline
Parameter & Simulated value & Ref. catalogue value & Unit \\\\
\hline \hline                       
General characteristics  & & &\\\hline
Random seed & 20251211082334 & & \\ 
Reference catalogue & & \citet{2015ApJS..219...12A} & \\
Number of galaxies (common footprint\tablefootnote{This number is measured in a chosen footprint for with both mock and reference catalogues present main data without border effects. The selected footprint limits are: $RA\in[140^{o}, 230^{o})$, $DEC\in[0^{o}, 50^{o})$ and $Dist.\in[100, 500)[h^{-1}Mpc]$}) & {${\num{275487}\pm\num{6700}}$} & {\num{274975}} & galaxies \\
R.A. range (R.A. max. - R.A. min.) & 120 & 152.05 & [\textsuperscript{o}]\\
DEC range (DEC max. - DEC min.) & 90 & 73.99 & [\textsuperscript{o}]\\
Mean density of galaxies (common footprint) & $0.0057\pm0.0001$ & 0.005 & galaxies $\times$ (h\textsuperscript{-1}Mpc)\textsuperscript{-3}\\\\
\hline\hline
Clusters &  &   \\\hline
Reference catalogue & & \citet{2017AA...602A.100T} & \\
Number of cluster galaxies & ${\num{16406}\pm\num{832}}$ & \num{16419} & galaxies\\
Number of clusters & ${\num{333}\pm\num{17}}$ & 322 & clusters \\
Min. nº of galaxies within a cluster & 30 & 30 & galaxies\\
Max. nº of galaxies within a cluster & $\num{291}$ & 254 & galaxies\\
Density of galaxies within clusters & ${\num{0.27}\pm\num{0.01}}$ & & galaxies $\times$ (h\textsuperscript{-1}Mpc)\textsuperscript{-3}\\
Clusters connectivity (node degree) & ${4}$ & & {filaments / node}\\\\
\hline\hline
Walls \& filaments &   \\
\hline
Number of wall and filament galaxies & ${\num{403150}\pm\num{7643}}$ & & galaxies  \\
Number of walls & ${\num{15022}\pm\num{569}}$ & & walls \\
Density of galaxies within walls and filaments & ${\num{0.023}\pm\num{0.001}}$ & & galaxies $\times$ (h\textsuperscript{-1}Mpc)\textsuperscript{-3}\\
Walls surface

    & ${\num{516.34}\pm\num{504.89}}$ & & (h\textsuperscript{-1}Mpc)\textsuperscript{2}\\
Filaments length

    & ${\num{31.18}\pm\num{27.16}}$ & & h\textsuperscript{-1}Mpc\\\\
\hline\hline
Voids &   \\
\hline
Reference catalogue & &\citet{2012MNRAS.421..926P} & \\
Number of void galaxies & ${\num{79387}\pm\num{3272}}$ & \num{79947} & galaxies \\
Number of voids & ${\num{1060}\pm\num{37}}$ & \num{1054} & voids\\
Density of galaxies within voids & $0.0010\pm0.0001$ & & galaxies $\times$ (h\textsuperscript{-1}Mpc)\textsuperscript{-3}\\
Median effective radius & ${\num{28.61}\pm\num{11.28}}$ & 17-25 & h\textsuperscript{-1}Mpc\\
Number of adjacent walls & ${\num{14.28}\pm\num{7.15}}$ & & walls / void\\
Number of adjacent filaments & ${\num{39.78}\pm\num{10.89}}$ & & filaments / void\\
Number of adjacent nodes & ${\num{26.95}\pm\num{6.94}}$ & & nodes / void\\
Void shape
    \tablefootnote{Void shape was computed for each void as the fraction between the sum of the surface of its surrounding walls and the area of a sphere whose $r=R_{eff}$.}
    & ${\num{1.28}\pm\num{0.35}}$ & & dimensionless\\
\\\hline
\hline
\end{tabular}
\end{table}

\clearpage
\twocolumn
\section{Density of galaxies along cluster radius}
\label{ex:cluster_shells}
Spatial distribution of galaxies along the cluster radius is highly
inhomogeneous. For instance, consider a cluster with $n$ galaxies, where its
furthest galaxy is at ${\sim}2$ Mpc from its geometrical centre, and the spatial
distribution of galaxies along its radius follows a Gaussian distribution,
centred in ${r=\mu=0}$ Mpc and with a standard deviation of
$\sigma={{\sim}\frac{2}{3}}$ Mpc. In this example, three slices (or `shells')
were studied according to several ranges of radial distances in the cluster: i)
from the centre ($\rm r_0$=0 Mpc) to $1\sigma$ ($\rm r_1$=0.66 Mpc); ii) from
$1\sigma$ ($\rm r_1$=0.66 Mpc) to $2\sigma$ ($\rm r_2$=1.32 Mpc); and iii) from
$2\sigma$ ($\rm r_2$=1.32 Mpc) to $3\sigma$ ($\rm r_3$=2 Mpc). In
Table~\ref{table:2_1_cluster}, the inner and outer radii (columns 2 and 3
respectively) of each studied shell, its galaxy proportion (column 4), its
volume (column 5) and its galaxy density (column 6) are shown. In the last row
of Table~\ref{table:2_1_cluster} the total number of galaxies inside the cluster
and its total volume is considered, giving us a mean density that matches with
the previously calculated one. The densities within the shells vary by several
orders of magnitude, making it inappropriate to determine or apply an average
galaxy density for clusters given their spatial distribution.

\begin{table}[h]
\caption{Galaxy densities versus radial slices for cluster structures.}
\label{table:2_1_cluster}
\centering
\begin{tabular}{rrrrrr}
\hline
\hline
Shell & Inner  & Outer   & Galaxies & Volume & Density \\
      & radius & radius  &          &        &         \\
& Mpc & Mpc & \% & [h\textsuperscript{-1}Mpc]\textsuperscript{3} & $[\frac{\textrm{galaxies}}{\textrm{(h\textsuperscript{-1}Mpc)\textsuperscript{3}}}]$\\
\hline
\hline
i   & 0.00 & 0.66 & 68.3 & 1.20  & 0.570$n$\\
ii  & 0.66 & 1.32 & 27.2 & 8.43  & 0.032$n$\\
iii & 1.32 & 2.00 & 4.2  & 23.88 & 0.0017$n$\\
\hline
Total & 0.00 & 2.00 & 99.7 & 33.51 & 0.0298$n$\\
\hline
\hline
\end{tabular}
\tablefoot{ $n$
 represents the number of galaxies within the cluster. In the example,
 $n=\textrm{Total mass of cluster}/\textrm{Mean galaxy
 mass}=10^{14}M_{\odot}/10^{11}M_{\odot}=\num{1000}\,[\textrm{galaxies}]$.}
\end{table}

\clearpage

\twocolumn
\section{Additional figures}

\begin{figure}
        \begin{center}
                \includegraphics[width=\hsize]{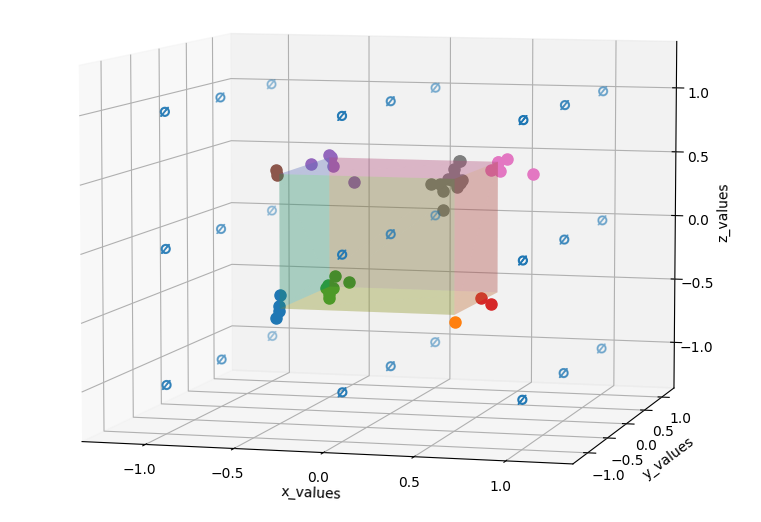}
                \caption[Voronoi 3D generated central cube region.]{Voronoi 3D generated central cube region. The random cluster galaxies grow around the
                vertices, acting as the centroid of each cluster. The empty symbols are the simulated seeds of voids. The points represent simulated cluster galaxies coloured depending on their belonging cluster.
        Right panel: }
                \label{fig:2_1_4_cluster_vertex}
        \end{center}
\end{figure}

\begin{figure}
        \begin{center}
        \includegraphics[width=\hsize]{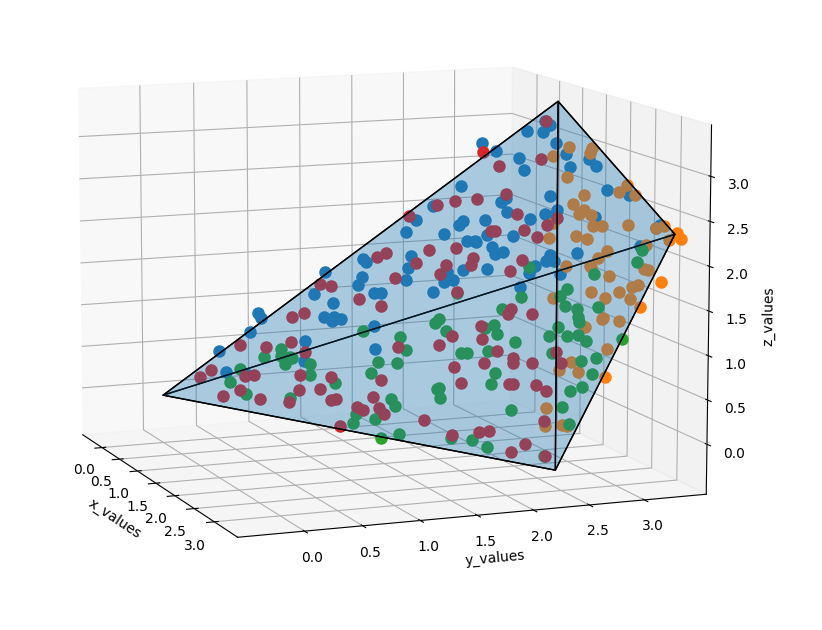}
        \caption{Example of a region from a Voronoi 3D 
        tessellation with its faces populated with random points representing wall and filament mock 
        galaxies. Extracted planes from the Voronoi tessellation, interpreted as cosmic wall (blue faces), extracted edges from the Voronoi tessellation, interpreted as cosmic filaments (black lines), and simulated wall galaxies coloured depending on their belonging wall (coloured points).}
        \label{fig:6_wall}
        \end{center}
\end{figure}

\begin{figure}
    \begin{center}
    \includegraphics[width=\hsize]{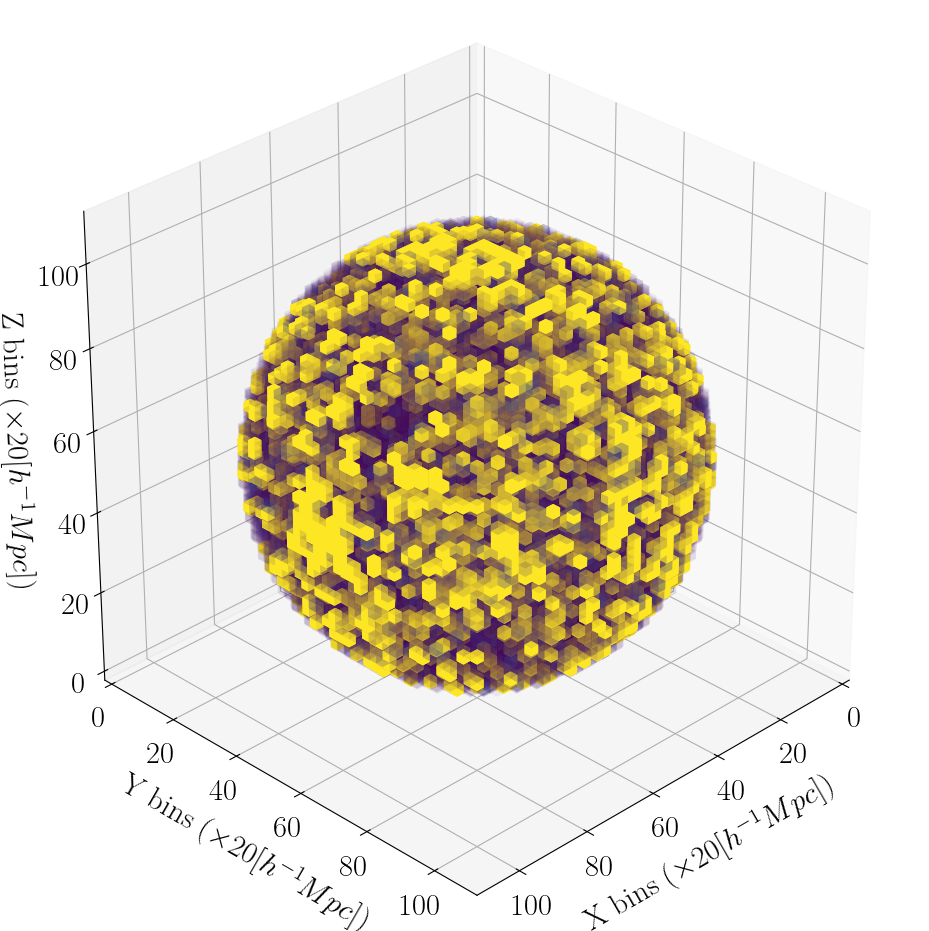}
    \end{center}
    \caption[Void galaxies probability map]
    {3D example of a void galaxies probability map, generated by
    inverting a 3D histogram performed over the galaxies from other structures.
    Yellowish and violetish voxels represents a spatial region with the highest and lowest probability of
    generating void galaxies, respectively. The size of the voxels
    was oversized to 20\~[h\textsuperscript{-1}Mpc]\textsuperscript{3} for a clearer
    representation.}
    \label{fig:inverted_3d_hist}
\end{figure}

\begin{figure}
    \begin{center}
    \includegraphics[width=\hsize]{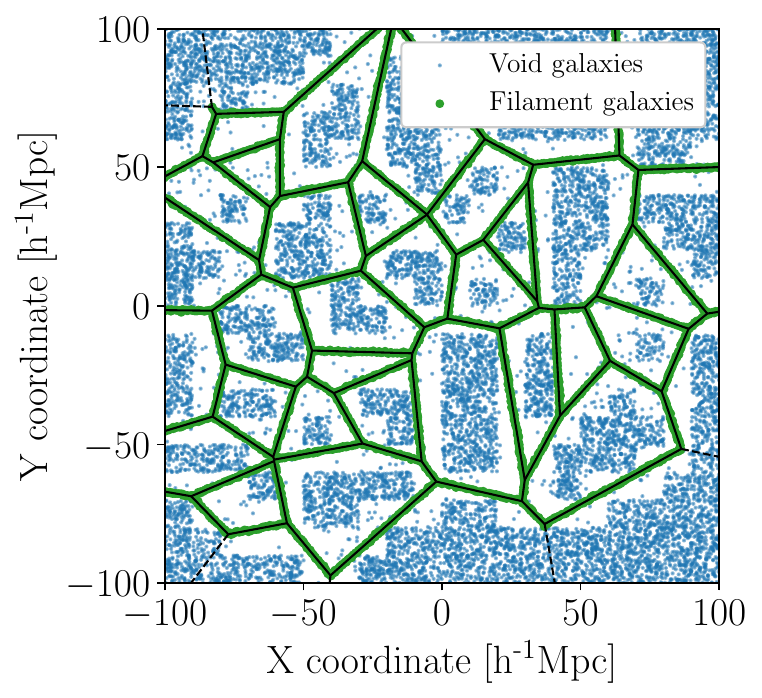}
    \end{center}
    \caption[Populating under-dense areas with void galaxies]
    {2D example of the under-dense areas in the void galaxies generation
    method. The figure shows a Voronoi tessellation of a surface. Black
    lines represents the filaments which were populated with galaxies
    (green dots). Blue dots represents generated
    uniformly distributed void galaxies. The threshold was exaggerated to
    highlight how the algorithm skips areas previously populated with galaxies
    from other structures. }
    \label{fig:2_1_5_void}
\end{figure}

\begin{figure*}
    \begin{center}
    \includegraphics[width=.85\hsize]{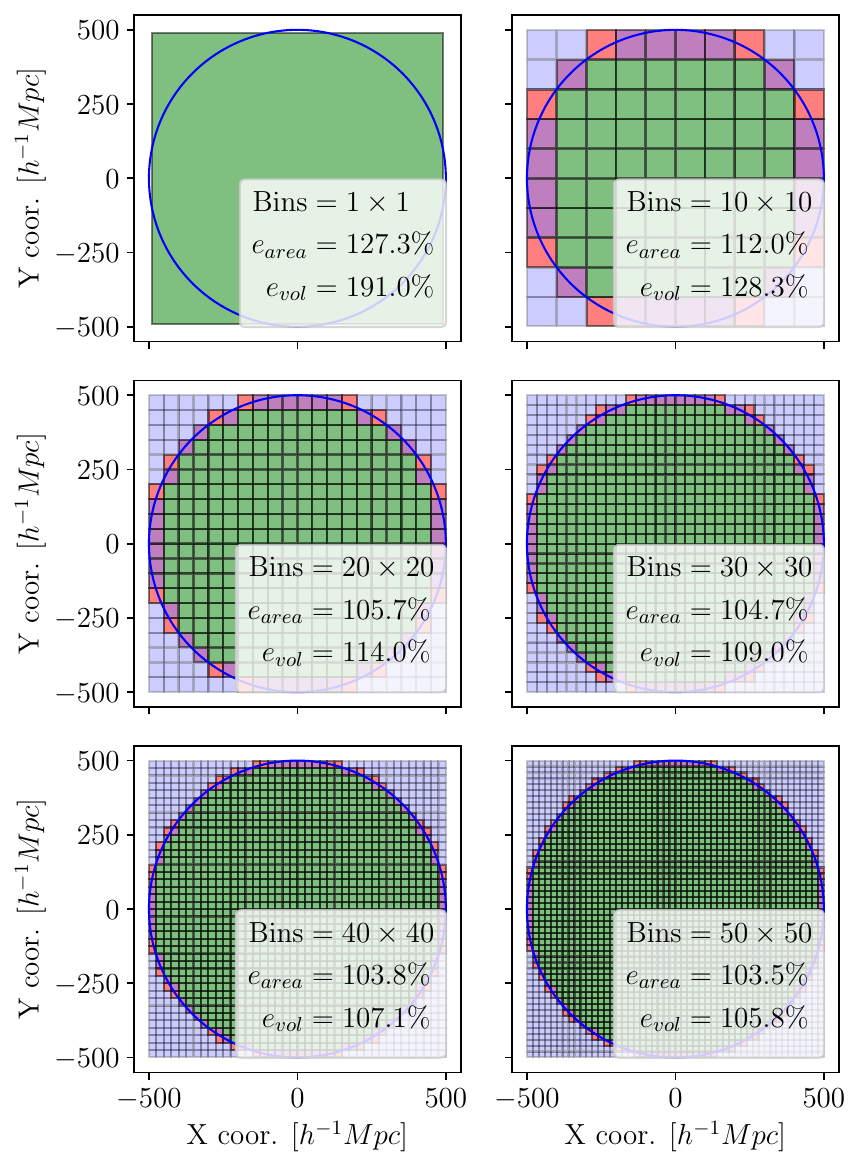}
    \end{center}
    \caption[Effect of sampling the sphere volume using voxels]
    {Effect of sampling the sphere volume using different voxel sizes. The figure shows a
    2D representation of the simulated spherical universe (blue
    circle). The voxel volume method will divide the space in voxels,
    represented as coloured squares. Those voxels which their centres are
    further than the simulated universe radius (500\,$[h^{-1}Mpc]$ in this
    example) will not be processed, represented in light blue. The remaining voxels will be
    taken into account to compute the volume of the structures. While inner
    voxels present no problem (green ones), frontier voxels are adding more
    volume erratically because cubes cannot precisely fit  the surface of a sphere.
    While purple voxels add small errors, red voxels are adding the expected
    volume twice or more for that zone. This 2D example shows how the
    proportion between the computed and theoretical areas (represented by
    $e_{area}$) decreases proportionally with the bin size. In 3D, it occurs in the same way with computed and theoretical
    volumes, represented by $e_{vol}$.}
    \label{fig:samplingvoxels}
\end{figure*}

\begin{figure*}
        \begin{center}
                \includegraphics[angle=0.0, width=0.90\hsize]{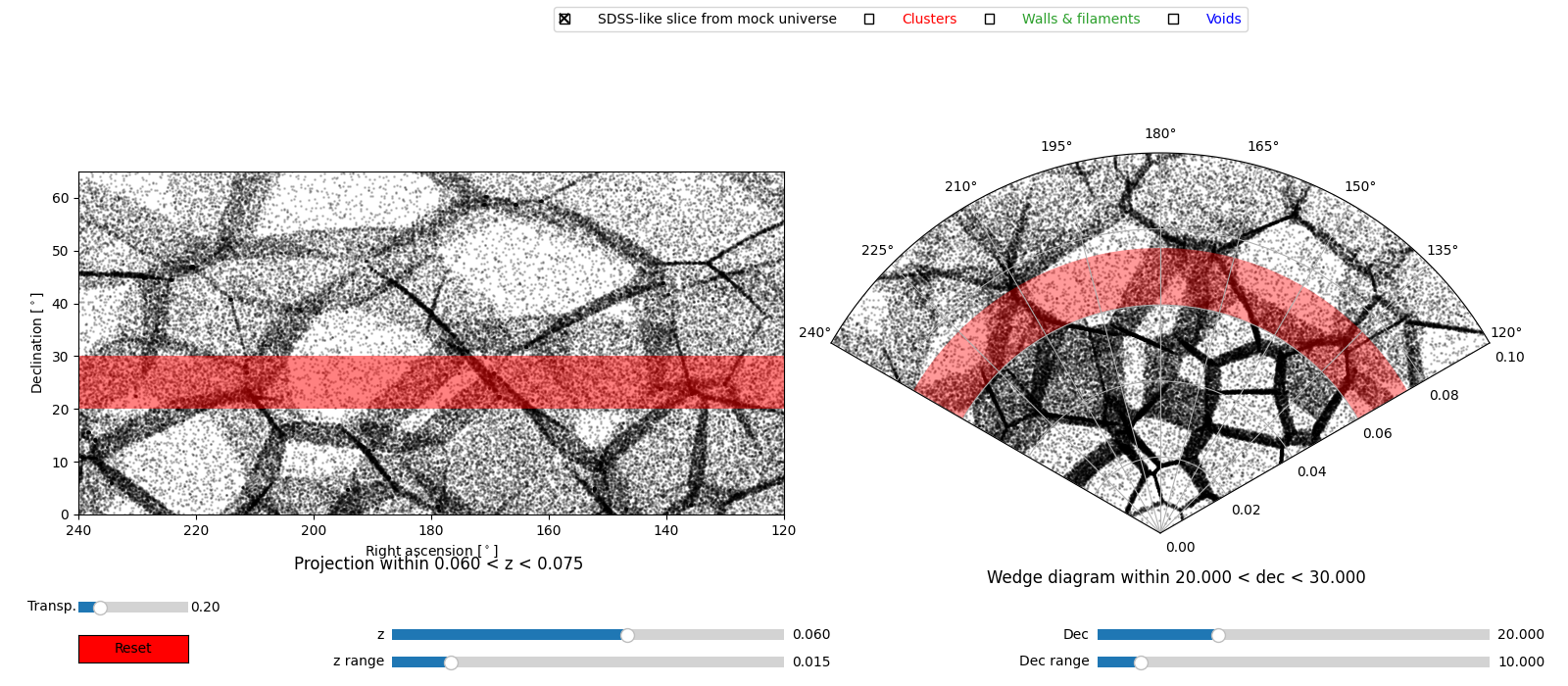}
        \includegraphics[angle=0.0, width=0.90\hsize]{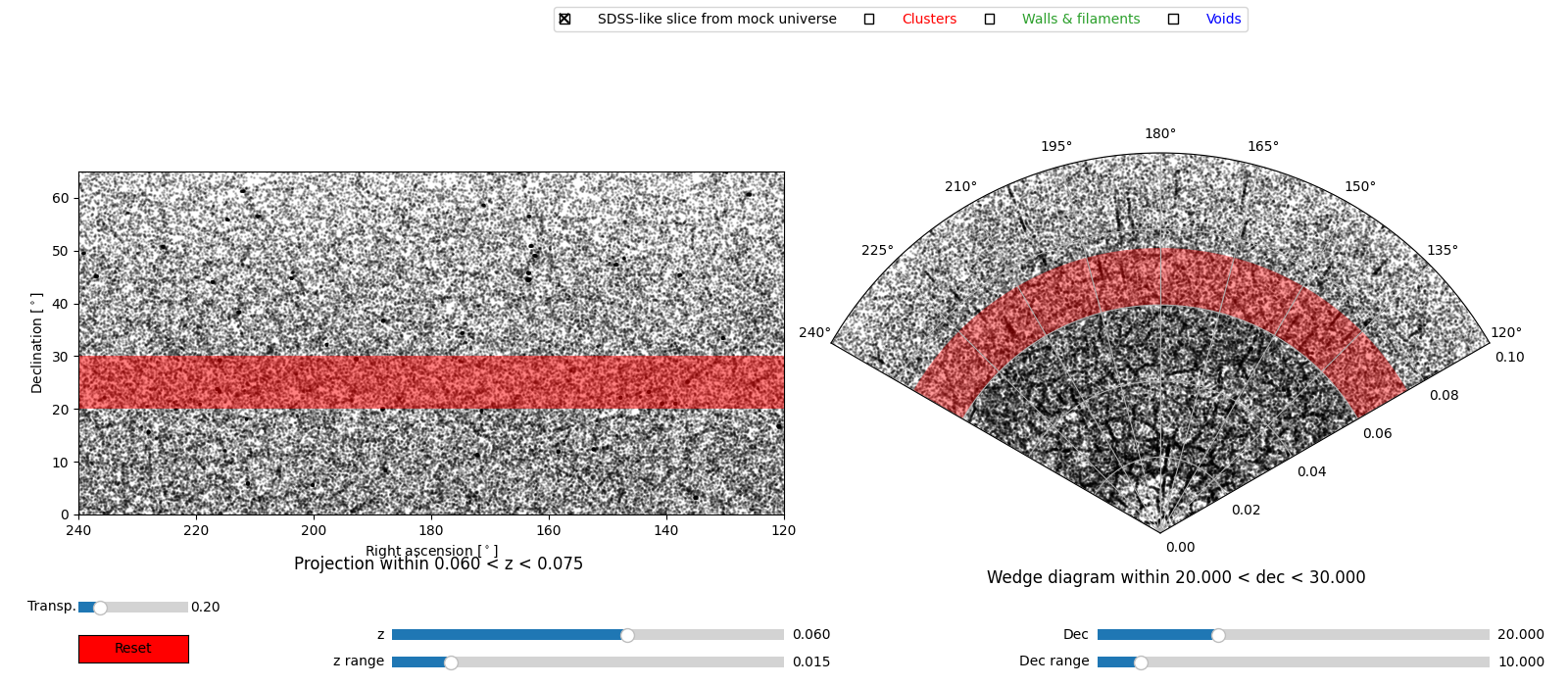}
                \caption{SDSS-like slice II from two randomly generated mock universes with ten times fewer (upper panel) and ten times more (lower panel) voids than in default configuration. In both panels, the red stripes represent the range of redshift and declination shown in the other sub-panel.}
                \label{fig:mock_100_10000}
        \end{center}
\end{figure*}

\end{document}